%Paper: hep-ph/9405375
%From: kugo@gauge.scphys.kyoto-u.ac.jp
%Date: Wed, 25 May 94 10:23:16 +0900

%%%%%%%%%%%%%%%%%%%%%%%%%%%%%%%%%%%%%%%%%%%%%%%%%%%%%%%%%%%%%%%%%
%%                                                             %%  
%% DEFINING THE NAMBU--JONA-LASINIO MODEL BY HIGHER DERIVATIVE %%
%% KINETIC TERM,   by Takashi HAMAZAKI and Taichiro KUGO       %%
%%                                                             %%
%%   Unpack the uuencoded eps figures appended to this file.   %%
%%   Please search for 'CUT HERE'.                             %%
%%%%%%%%%%%%%%%%%%%%%%%%%%%%%%%%%%%%%%%%%%%%%%%%%%%%%%%%%%%%%%%%%

\input phyzzx

%%%%%%%%%%%%%%%%%%%% for figures %%%%%%%%%%%%%%%%%%%%%%%%%%%%%%
%%% Examining the system automatically %%%
\newif\iffigureexists
\newif\ifepsfloaded
\openin 1 epsf
\ifeof 1 \epsfloadedfalse \else \epsfloadedtrue \fi
\closein 1
\ifepsfloaded
    \input epsf
\else
    \immediate\write20{>Warning:
         No epsf.tex --- cannot embed Figures!!}
\fi
\def\checkex#1 {\relax
    \ifepsfloaded \openin 1 #1
        \ifeof 1 \figureexistsfalse
        \else \figureexiststrue
        \fi \closein 1
    \else \figureexistsfalse
    \fi }
\ifepsfloaded
\checkex{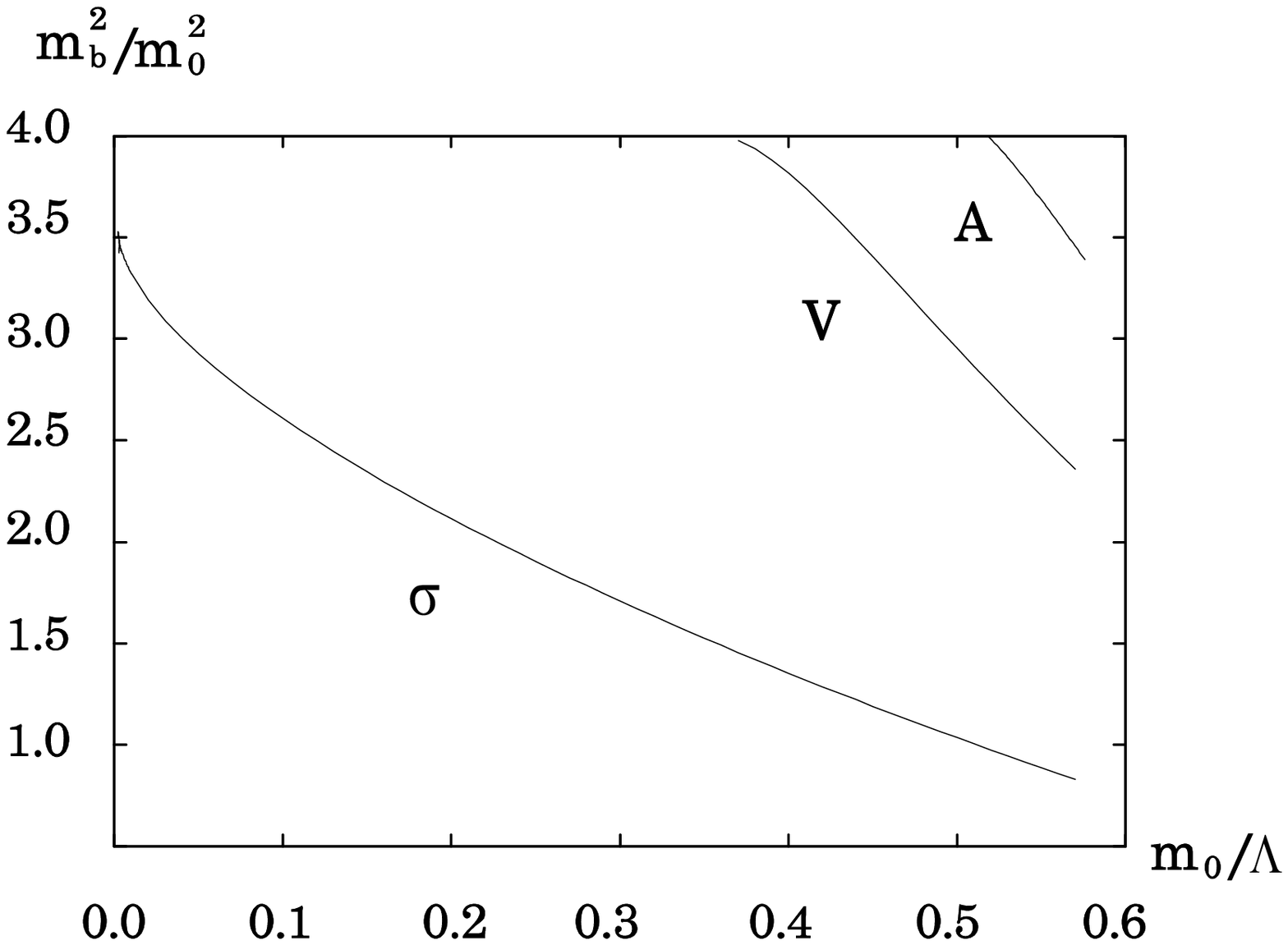}
    \iffigureexists \else
    \immediate\write20{>EPS files for the Figs. 1 -- 3 are packed
     in a uuecoded compressed tar file}
    \immediate\write20{>appended to this TeX file.}
    \immediate\write20{>You should unpack them and TeX again!!}
    \fi
\fi
%%%  Following \figinsert command is used like
%     << \figinsert{Fig#}
%                  {filename.eps}{xsize}
%                  {Here is a figure caption}. >>
%%%%%%%%%%%%%%%%%%%%%%%%%%%%%%%%%%%%%
\def\figinsert#1#2#3#4{
\checkex{#2}
\midinsert
\iffigureexists
    \immediate\write20{>Embedding an epsf Figure (#2).}
\epsfxsize=#3
\centerline{\epsfbox{#2}}
\else
    \immediate\write20{>Cannot embed a epsf Figure (#2).}
\fbox{Figure #1}
\fi
\captionbox#1{#4}
\endinsert }
\newbox\tempboxa
\newdimen\captionboxsubcount \captionboxsubcount=30pt
\newdimen\captionboxsub
\def\captionbox#1#2{
\captionboxsub=\hsize \advance\captionboxsub by -\captionboxsubcount
\setbox\tempboxa\hbox{ {\bf Fig.#1.}\ \ #2 }
\ifdim \wd\tempboxa >\captionboxsub
{ \narrower {\narrower \hangafter=1
  \hangindent=20pt \noindent{{\bf Fig.#1.}}\ \ #2\par} }
\else \hbox to\hsize{\hfil\box\tempboxa\hfil}
\fi  }
\def\fboxsub#1{\vbox{\hrule\hbox{%
\vrule\kern3pt\vbox{\kern3pt#1\kern3pt}\kern3pt\vrule}\hrule}}
\def\fbox#1{\setbox4=\vbox{\kern3pt\hbox{#1}\kern3pt}
\hfill\fboxsub{\box4}\hfill\ \par}
%
%%%%%%%%%%%%%%%%%%%%%% MACROS %%%%%%%%%%%%%%%%%%%%%%%%%%%%%%%%%

\def\calM{{\cal M}}
\def\calV{{\cal V}}
\def\calA{{\cal A}}
\def\dirac{{\rlap/ \partial}}
\def\pslash{{p \kern -5.2pt /}}
\def\qslash{{q \kern -5.2pt /}}
\def\kslash{{k \kern -5.2pt /}}
\def\Dslash{{D \kern -7.5pt /\,}}
\def\Vslash{{\calV \kern -7.5pt /\,}}
\def\Aslash{{\calA \kern -7pt / \,}}
\def\chiral{e^{i{\gamma_5}\theta}}
\def\-chiral{e^{-i{\gamma_5}\theta}}
\def\VL{{\bf L}}
\def\bpartial{{\buildrel \leftrightarrow \over \partial}}

\def\4mp{{4{m^2}-{p^2}}}

\def\Yukawa{{g_{\pi {\bar \psi} \psi}}}

\def\trans{{{\hat \delta}_a \phi}}
\def\NJL{Nambu--Jona-Lasinio}

\def\nf{n_{\rm f}}
\def\Nc{N_{\rm c}}
\def\R{{\rm R}}
\def\L{{\rm L}}

\def\V{{\rm V}}
\def\VZ{{\rm V0}}
\def\S{{\rm S}}
\def\calPR{{\cal P}_{\rm R}}
\def\calPL{{\cal P}_{\rm L}}
\def\tree{
  -{\Lambda^2\over2g_\S^2}\tr\big(\Sigma^\dagger \Sigma\big)
  +{\Lambda^2\over2g_\V^2} \tr\big(R^\mu R_\mu+L^\mu L_\mu\big) }
\def\FT{{\rm F.T.}}
\def\Tr{{\rm Tr}}
\def\Ln{{\rm Ln}}
\def\mymatrix#1#2#3#4#5#6#7#8#9{\left( \matrix{#1  &  #2  & #3 \cr
                #4  &  #5  & #6 \cr #7  &  #8 & #9 \cr} \right) }
\def\myvector#1#2#3{\pmatrix{#1 \cr #2  \cr #3 \cr}}
\def\Mij{\overline{m}_{ij}^2}
\def\reg#1{[#1]_\Lambda}
\def\AVF{{\cal A}^\mu(k,k+p)}
\def\Mpgg{{\cal M}_{\pi_0\rightarrow\gamma\gamma}}

%%%%%%%%%%%%%%%%%%%%% from Particle.tex %%%%%%%%%%%%%%%%%%
\def\ee{\eqno\eq }

%square (in math mode)
%
\def\sqr#1#2{{\vcenter{\hrule height.#2pt
      \hbox{\vrule width.#2pt height#1pt \kern#1pt
          \vrule width.#2pt}
      \hrule height.#2pt}}}
\def\square{{\mathchoice{\sqr84}{\sqr84}{\sqr{5.0}3}{\sqr{3.5}3}}\,}
%
%%%%%%%%%%%%%% Macros for the Title Page
\catcode`@=11
\newtoks\KUNS
\newtoks\HETH
\newtoks\monthyear
\Pubnum={KUNS~\the\KUNS\cr HE(TH)~\the\HETH\cr hep-ph/9405375}
\monthyear={\monthname,\ \number\year}
\def\p@bblock{\begingroup \tabskip=\hsize minus \hsize
 \baselineskip=1.5\ht\strutbox \topspace-2\baselineskip
 \halign to\hsize{\strut ##\hfil\tabskip=0pt\crcr
 \the\Pubnum\cr \the\monthyear\cr }\endgroup}
\def\bftitlestyle#1{\par\begingroup \titleparagraphs
 \iftwelv@\fourteenpoint\else\twelvepoint\fi
 \noindent {\bf #1}\par\endgroup }
\def\title#1{\vskip\frontpageskip \bftitlestyle{#1} \vskip\headskip }
%
%%%%%%%%%%%%%% Address of Kyoto University
%
\def\Kyoto{\address{Department of Physics,~Kyoto University \break
 Kyoto~606-01,~JAPAN}}
%
%%%%%%%%%%%%%% Style of page numbers
%
\paperfootline={\hss\iffrontpage\else\ifp@genum%
 \tenrm --\thinspace\folio\thinspace --\hss\fi\fi}
\footline=\paperfootline
%
%%%%%%%%%%%%%% Macros for References
%
% Let's make \refmark larger

%
%Redefinition of \journal and new macros
\def\journal#1&#2(#3){\begingroup \let\journal=\dummyj@urnal
 \unskip, \sl #1\unskip~\bf\ignorespaces #2\rm
 (\afterassignment\j@ur \count255=#3) \endgroup\ignorespaces }
\def\andjournal#1&#2(#3){\begingroup \let\journal=\dummyj@urnal
 \sl #1\unskip~\bf\ignorespaces #2\rm
 (\afterassignment\j@ur \count255=#3) \endgroup\ignorespaces }
\def\andvol&#1(#2){\begingroup \let\journal=\dummyj@urnal
 \bf\ignorespaces #1\rm
 (\afterassignment\j@ur \count255=#2) \endgroup\ignorespaces }

\def\PR{Phys.~Rev. }

\def\PL{Phys.~Lett. }
\def\PTP{Prog.~Theor.~Phys. }

%
%%%%%%%%%%%%%% Macro for Acknowledgements
\def\acknowledge{\par\penalty-100\medskip \spacecheck\sectionminspace
 \line{\hfil ACKNOWLEDGEMENTS\hfil}\nobreak\vskip\headskip }
\catcode`@=12
%%%%%%%%%%%%%%%%%%%%%%%%%%%%%%%%%%%%%%%%%%%%%%%%%%%%%%%%
%%%%%%%%%%%%%%%%%%% REFERENCES %%%%%%%%%%%%%%%%%%%%%%%%%
\REF\NJ{Y.~Nambu and G.~Jona-Lasinio \journal \PR &122 (61) 345;
\andvol &124 (61) 246.}
\REF\HK{For an excellent review of NJL model and the
application to QCD, see \nextline
T.~Hatsuda and T.~Kunihiro, Tsukuba preprint UTHEP-270, to appear
in {\sl Physics Report} (1994).}
\REF\MR{See many references cited in Ref.2).}
\REF\BH{W.A.~Bardeen and C.~Hill \journal \PR &D49 (94) 409.}
\REF\PU{A.~Pais and G.E.~Uhlenbeck \journal \PR &79(50) 145.}
\REF\Nakanishi{N.~Nakanishi, {\sl \PTP} Supplement  No.51 (1972) 1.}
\REF\Slavnov{A.A.~Slavnov \journal Teor. Mat. Fiz. &33 (77) 210.}
\REF\Maekawa{N.~Maekawa, Master Thesis (in Japanese),
Kyoto University, 1990;
in Proc. 1989 Nagoya Workshop on {\it Dynamical Symmetry Breaking},
ed. by T.~Muta and K.~Yamawaki, p63; \nextline
K.~Suehiro, in Proc. of 1990 Nagoya International Workshop on
{\it Strong Coupling Gauge Theories and Beyond}, ed. by
T.~Muta and K.~Yamawaki, (World Scientific, 1991) p93; \nextline
C.T.~Hill, {\it ibid}, p37.}
\REF\MTY{V.A.~Miransky, M.~Tanabashi and K.~Yamawaki \journal \PL
&B221 (89) 177; \andjournal Mod. Phys. Lett. &A4 (89) 1043;
\nextline
Y.~Nambu, preprint FPI 89-39, 1989; in Proc. 1988 Nagoya International
Workshop on {\it New Trends in Strong Coupling Gauge Theories}, ed. by
M.~Bando, T.~Muta and K.~Yamawaki (World Scientific, Singapore, 1989);
\nextline
W.A.~Bardeen, C.T.~Hill and M.~Lindner \journal \PR &D41 (90) 1647.}
\REF\Jackiw{R.~Jackiw, in {\it Current Algebra and Anomaly},
ed. by S.~Treiman, R.~Jackiw, B.~Zumino and E.~Witten
(World Scientific, Singapore, 1985).}

%%%%%%%%%%%%%%%%% TITLE PAGE %%%%%%%%%%%%%%%%%%%%%%%%%
%\nopubblock    %This command deletes the preprint number block.
\KUNS={1263}    %KUNS number.
\HETH={94/07}   %HE/TH number.
\monthyear={May, 1994} %If you want to fix the date.

\titlepage

\title{ Defining the Nambu--Jona-Lasinio Model by  Higher Derivative
Kinetic Term}

\author{Takashi HAMAZAKI and Taichiro KUGO}

\Kyoto  %or \KYOTO (Address of Dept. of Phys., Kyoto Univ.)

\abstract{
Usual treatment of the Nambu--Jona-Lasinio (NJL) model
using loop momentum cutoff
suffers from ambiguities in choosing the loop momenta to be cut
off and violation of (external) gauge invariance.
We define the NJL model from the starting Lagrangian
level by using a higher derivative fermion kinetic term with
a cutoff parameter $\Lambda $.
This definition is free from such ambiguities and manifestly keeps
the chiral symmetry as well as the gauge invariance.
Quantization of this higher derivative system, current operators
and calculational method are discussed in some detail.
Calculating the pion decay constant and $\pi ^0\rightarrow 2\gamma $
decay amplitude,
we explicitly demonstrate that the low energy theorem holds.
It is observed that the NJL mass relation $m_\sigma = 2 m_0$
between the $\sigma$ meson and fermion masses no longer holds in this
model.
We also present a simplified calculational method which is valid
when the cutoff parameter $\Lambda $ is much larger than the energy
scale
of physics.
}

\endpage

\chapter{Introduction}

The \NJL\ (NJL) model\refmark{\NJ,\HK} is not defined without
an ultraviolet cutoff $\Lambda $.
Usually this cutoff is introduced at the loop graph level.
For instance, in the most popular treatment,
the fermion one-loop integral is first rewritten using the Feynman
parameter formula and the ultraviolet cutoff is made
for the loop momentum variable with which the denominator of the
integrand becomes an even function. Another treatment, which was
adopted in the original NJL paper\rlap,\refmark{\NJ}
utilizes the dispersion
relation. In either treatment, the cutoff is made graph by graph
(or for each Green function separately).
This implies that the theory is {\it not} defined at the starting
Lagrangian level. This is very unsatisfactory.

Other serious problems in the usual treatment
are the ambiguity in introducing the cutoff
and the consistency with symmetries.
The values of the divergent loop diagrams depend on the choice of
the loop momentum variables for which we make the ultraviolet cutoff.
We should specify the choice procedure unambiguously.
Moreover the procedure
has to be shown to be consistent with the chiral symmetry at least.
To show this consistency would not be an easy task if the
cutoff is introduced graph by graph.
Furthermore if the system is coupled to external gauge fields
the cutoff procedure should also be consistent with the gauge
invariance. It is indeed very difficult to satisfy the gauge invariance
if we adopt such a simple cutoff for the loop momentum variables.

A best example for the
last problem is given by vector 2-point functions.
Consider a NJL model which contains 4-fermi interaction in the
vector channel and couples to the external photon field $A_\mu $.
If one rewrite the 4-fermi interaction term by introducing
a vector auxiliary field $V_\mu $, the fermion kinetic term gets
to take the form
$$
\bar\psi i\gamma ^\mu (\partial _\mu -iV_\mu -ieA_\mu )\psi .
\ee
$$
Note the pararellism between the vector field $V_\mu $ and the
photon field $A_\mu $. In calculating the fermion one-loop diagrams,
there is no difference between them; namely, exactly the same Feynman
diagrams appear for the vector-vector, vector-photon and photon-photon
2-point functions. If we adopt a loop momentum cutoff
to the diagram, we would obtain gauge non-invariant
function not proportional to $g_{\mu \nu }p^2-p^\mu p^\nu $, which may
be a good result for the vector-vector function but is clearly
unacceptable for the photon-photon and vector-photon functions.
To achieve the gauge invariance for the latter, one sometimes throw
away
the gauge non-invariant piece, quadratic mass term $\propto g_{\mu \nu
}\Lambda ^2$,
by hand. This might be a correct procedure. But if so, then, should
we do the same also for the vector-vector case?

We propose in this paper to define the NJL model
by using a higher derivative kinetic term for the fermion.
We replace the usual fermion kinetic term
by the following higher derivative one:
$$
\bar\psi  i\dirac \psi  \ \Rightarrow \
\bar\psi  i\dirac \big( 1+{\dirac \dirac\over \Lambda ^2} \big)\psi \ .
\ee
$$
Here the parameter $\Lambda ^2$ plays the role of an ultraviolet
cutoff.
Note that this effective cutoff is
made on each {\it fermion propagator} but not on
each loop momentum, and that any loop diagrams are now well-defined and
independent of the choice of the loop momenta.
It should be emphasized that this defines the NJL model
already at the starting {\it Lagrangian level}. It is manifest that
this higher derivative definition keeps
the important chiral symmetry of the system.  Moreover if we switch on
external gauge interactions (such as weak-electromagnetic ones),
we can keep the gauge-invariance also by simply replacing
each derivative factor $\dirac$ by a covariant one $\Dslash$.

As a mere regularization method there is  dimensional
regularization or zeta function regularization which respects
the gauge invariance.
But what we need here is not a mere regularization but a definition of
the NJL model by a `regularization' which is not removed forever.
If we define the NJL model by dimensional regularization, the model
will be defined in $4-\epsilon $ dimensions with a certain $\epsilon $
fixed
and then the physical meaning would become very unclear. In our
definition
by higher derivative, the cutoff parameter $\Lambda $ has a physical
meaning
as the energy scale which gives an upper limit of the applicability of
the model.

This paper is organized as follows.
In Sect.2 we set up the NJL model which we discuss in this paper.
Since the model is a higher derivative system rather different
from the canonical one, we discuss there
the problems of quantization, current operators and coupling to the
external gauge fields in detail. In Sect.3 we present a method for
reducing the calculations of Feynman diagrams in this system to
those of the usual first-order derivative fermion system. We calculate
some Green functions and note, in particular that the scalar meson
has a pole below the two fermion threshold. In Sect.4
$\pi ^0 \rightarrow  2\gamma $ amplitude is calculated and is
explicitly demonstrated
to be consistent with the low energy theorem. When the cutoff $\Lambda
$
is much larger than the energy scale we discuss, the calculation
can be made much simpler. This is shown in Sect.5. Sect.6 is devoted
to conclusion.
In the Appendix, we present a generalized Noether procedure of
constructing conserved current in a generic higher derivative system.

\chapter{ NJL Model Defined by Higher Derivative Kinetic Term}

\section{NJL Model and QCD-analogue Model}

We consider a fermion $\psi =(\psi _{if})$
which carries $SU(\Nc)$ color index $i$ and $SU(\nf)$ flavor index $f$.
The fundamental representation matrices are denoted by
$T^A \ (A=1,\cdots ,\Nc^2-1)$ for the color $SU(\Nc)$ and by
$\lambda ^a \ (a=1,\cdots ,\nf^2-1)$ for the flavor $SU(\nf)$,
respectively.
They are normalized by $\tr(T^AT^B)={1\over 2}\delta ^{AB}$,
$\tr(\lambda ^a\lambda ^b)=2\delta ^{ab}$.  We also use flavor singlet
matrix
$\lambda ^0\equiv \sqrt{2/\nf}\,{\bf 1}_{\nf}$
proportional to $\nf\times \nf$ unit matrix ${\bf 1}_{\nf}$.
The NJL model we consider in this paper is the following system
which possesses chiral $U(\nf)_{\rm R}\times U(\nf)_{\rm L}$ symmetry
up to the axial $U(1)_{\rm A}$ anomaly:
$$
\eqalign{
{\cal L} =\ &\bar\psi  i\dirac \big( 1+{\dirac \dirac\over \Lambda ^2}
\big)\psi  \cr
     &+{g_\S^2\over 4\Lambda ^2}\Big\{
        \big[ (\bar\psi {\lambda ^0}\psi )^2 + (\bar\psi i\gamma
_5{\lambda ^0}\psi )^2 \big]
        + \big[ (\bar\psi {\lambda ^a}\psi )^2
                        + (\bar\psi i\gamma _5{\lambda ^a}\psi )^2
\big]\Big\} \cr
     &-{g_\VZ^2\over 8\Lambda ^2}
      \big[ (\bar\psi \gamma _\mu {\lambda ^0}\psi )^2
        + (\bar\psi \gamma _\mu \gamma _5{\lambda ^0}\psi )^2 \big]
         -{g_\V^2\over 8\Lambda ^2}\big[(\bar\psi \gamma _\mu {\lambda
^a}\psi )^2
                        + (\bar\psi \gamma _\mu \gamma _5{\lambda
^a}\psi )^2 \big] \ .\cr
}\eqn\eqFFERMI
$$
Here all fermion bilinears are of {\it color-singlet}.
The fermion kinetic term is taken to be a higher derivative one as
explained in the Introduction. This fully defines the model from
the starting Lagrangian.

If we introduce auxiliary fields following the well-known procedure,
this Lagrangian can equivalently be rewritten into
$$
\eqalign{
{\cal L} &= {\bar \psi}\Big[i\dirac (1+{{\dirac
\dirac}\over{\Lambda^2}})-\calM\Big]\psi
   \tree \cr
   & \qquad \qquad
     +{\Lambda ^2\over 4}\left({1\over g_\VZ^2}-{1\over g_\V^2}\right)
     \left(\left[\tr(R_\mu \lambda ^0)\right]^2 +\left[\tr(L_\mu
\lambda ^0)\right]^2\right)
\ ,\cr
\noalign{\vskip0.5cm}
&\
\calM(x)\equiv  \Sigma(x){\cal P}_R
+{\Sigma^\dagger}(x){\cal P}_L
-{\gamma^\mu} {R_\mu}(x){\cal P}_R
-{\gamma^\mu} {L_\mu}(x){\cal P}_L \ .
\cr}\eqn\eqLAGRANGIAN
$$
with chiral projection operators $\calPR\equiv (1+\gamma_5)/2$ and
$\calPL\equiv (1-\gamma_5)/2$.  Eq.\eqLAGRANGIAN\  gives
equations of motion for the auxiliary fields as follows:
$$
\eqalign{
\Sigma &= -{g^{2}_\S \over\Lambda^2}
\left[\lambda ^0(\bar\psi _\R \lambda ^0\psi _\L )+\lambda ^a(\bar\psi
_\R \lambda ^a\psi _\L ) \right]\cr
 R_\mu &= -{g^{2}_\V\over 2\Lambda^2}
\left[{g_\VZ^2\over g_\V^2}\lambda ^0(\bar\psi _\R\gamma _\mu \lambda
^0\psi _\R)
        +\lambda ^a(\bar\psi _\R\gamma _\mu \lambda ^a\psi
_\R)\right]\cr
 L_\mu &= -{g^{2}_\V\over 2\Lambda^2}
\left[{g_\VZ^2\over g_\V^2}\lambda ^0(\bar\psi _\L\gamma _\mu \lambda
^0\psi _\L)
        +\lambda ^a(\bar\psi _\L\gamma _\mu \lambda ^a\psi
_\L)\right]\cr
}\eqn\eqEQMOT
$$
with $\psi _{\R,\L}\equiv {\cal P}_{\R,\L}\psi $.
Note that all the auxiliary fields are $\nf\times \nf$ flavor matrices;
$\Sigma$ is a complex matrix while $R^\mu,L^\mu$ are hermitian matrices.

The NJL model is often used as a model simulating
QCD\rlap.\refmark{\MR}
If we consider a single gluon exchange,
it may effectively be expressed by the
following four-fermion interaction:\refmark{\BH}
$$
{\cal L}^{\rm int}_{\rm QCD\hbox{-}analogue}=
-{{g^{2}}\over{\Lambda^{2}}}(\bar \psi \gamma_\mu
{T^A} \psi)(\bar \psi \gamma^\mu{T^A} \psi ) \ ,
\eqn\eqQCD
$$
where $\Lambda $ is a (suitable) characteristic energy scale of QCD and
$g$ is the color gauge coupling constant. If we perform the Fierz
transformation and keep only the leading terms in $1/\Nc$,
this four-fermi interaction \eqQCD\ can be
rewritten into the same form as that in Eq.\eqFFERMI\ with
identification
$$
g_\S^2=g_\V^2=g_\VZ^2=g^2\ .
\eqn\eqQCDANALOGUE
$$
( If we also keep non-leading terms in $1/\Nc$, only the flavor-singlet
vector four-fermi coupling is replaced by
$g_\VZ^2=\left(1-{2\nf/\Nc}\right)g^2$. )
We refer to this four-fermi interaction system
with coupling relations \eqQCDANALOGUE\ as `QCD-analogue NJL
model'.

\section{Quantization}

Let us now consider the quantization of this higher derivative system.
The procedure has long been known since Pais and
Uhlenbeck\rlap,\refmark{\PU} and
we here follow the procedure by Nakanishi\rlap.\refmark{\Nakanishi}
Consider generally a system which contains higher derivatives only
in the kinetic term as follows:
$$
{\cal L}= \bar\psi f(i\dirac)\psi +{\cal L}_{\rm int}(\psi ,\bar\psi ),
\eqn\eqHDL
$$
where $f(x)$ is a polynomial of the form
$$
f(x)= a \prod _{j=0}^n (x-m_j) \qquad \qquad (m_j\not=m_k \ {\rm for}\
j\not=k)
\ee
$$
and the interaction part ${\cal L}_{\rm int}$ is assumed to contain no
derivatives.
The well-known partial fraction formula
$1/f(x)=\sum _j[f'(m_j)(x-m_j)]^{-1}$ leads to an identity
$$
\sum _{j=0}^n a\eta _j\big[\prod _{k\not=j}(x-m_k)\big] = 1
\quad \qquad \big( \eta _j={1\over f'(m_j)} \big) \ .
\ee
$$
Using this, we can decompose the fermion field $\psi $ as
$$
\psi = \sum _{j=0}^n \psi _j \, \qquad \ {\rm where}
\qquad \psi _j\equiv  a \eta _j\big[\prod
_{k\not=j}(i\dirac-m_k)\big]\psi \ ,
\eqn\eqCOMPONENT
$$
and then, by noting
$f(i\dirac)\psi =\eta _j^{-1}(i\dirac-m_j)\psi _j$ independently of
$j$,
the higher derivative Lagrangian \eqHDL\ is seen to be rewritten into
$$
{\cal L} = \sum _{j=0}^n\eta _j^{-1}\bar\psi _j(i\dirac-m_j)\psi _j
   + {\cal L}_{\rm int}\left(\psi \!=\sum \nolimits_j\psi _j,\
                       \bar\psi \!=\sum \nolimits_j\bar\psi _j\right) \
{}.
\eqn\eqDECOMPOSEDL
$$
It is now clear in this form that the original higher derivative system
is equivalent to a system consisting of ordinary
(positive or negative metric\foot{
If we arrange the roots $m_j$ in order of their values,
the weights $\eta _j=1/f'(m_j)$
take alternating signs and hence $\psi _j$ become
of positive and negative metric alternatingly.
}) fermion fields $\psi _j$ with mass $m_j$,
to which the usual canonical quantization procedure is applicable.
Then clearly the free propagator of the $j$-th fermion is given by
$i\eta _j/(\pslash-m_j)$ and therefore that of  the original field
$\psi $ is found to be given by
$$
\eqalign{
\FT\VEV{{\rm T}\psi \bar\psi }
&= \sum _{j=0}^n\FT\VEV{{\rm T}\psi _j\bar\psi _j} \cr
&= \sum _{j=0}^n\eta _j{i\over \pslash-m_j}
= {i\over  a \prod _{j=0}^n(\pslash-m_j)}
= {i\over f(\pslash)} \ .
\cr}\eqn\eqDECOMPOSITION
$$
Namely,
the naive expectation that the propagator of $\psi $
is given by the inverse
of the kinetic term is correct in this case.
Conversely, if we take it for granted that the propagator of $\psi $ is
given by $i/f(\pslash)$, we can start from it and decompose the
$\psi $-propagator into partial fractions
$\sum _{j=0}^n i\eta _j/(\pslash-m_j) $ in every Feynman diagram.
Then it is easy to see that the theory is equivalent to
the above system \eqDECOMPOSEDL\ consisting of
$(n+1)$-fermions $\psi _j$ with mass $m_j$.

We can apply this general procedure to our NJL system \eqLAGRANGIAN\
in various ways depending on which part we regard as the free kinetic
term ${\cal L}_{\rm free}\equiv \bar\psi f(i\dirac)\psi $. We now
discuss two ways,
separately.

\subsection{A picture}

The simplest way, which we call `A picture',
is to take the original kinetic term
$\bar\psi  i\dirac \big( 1+{\dirac \dirac / \Lambda ^2} \big)\psi $
as ${\cal L}_{\rm free}$. Then $f(x)=x(1-x/\Lambda )(1+x/\Lambda )$
and it has three roots $m_j= 0, \Lambda , -\Lambda $ with $\eta _j= 1,
-1/2, -1/2$.
Accordingly, the original fermion field $\psi $ is decomposed into
$\psi = \psi _0 + \psi _\Lambda  + \psi _{-\Lambda }$ with $\psi
_{0,\pm \Lambda }$ denoting the
component fermions with masses $m_j= 0, \Lambda , -\Lambda $,
respectively,
and the free part lagrangian is written in the form:
$$
\eqalign{
{\cal L}^{(0)}_{\rm free} &\equiv
\bar\psi  i\dirac \big( 1+{\dirac \dirac\over \Lambda ^2} \big)\psi  \cr
&= \bar \psi_0 i\dirac \psi_0
-{2}\bar \psi_{\Lambda}(i\dirac-\Lambda)\psi_\Lambda
-{2}\bar \psi_{-\Lambda}(i\dirac+\Lambda)\psi_{-\Lambda} \ .\cr
}\eqn\eqFREE
$$
This A picture hence corresponds to the following decomposition
of the fermion propagator:
$$
{1\over -\pslash(1-{p^2\over \Lambda ^2})} =
{1\over -\pslash} -{1\over 2}{1\over \Lambda -\pslash}
 -{1\over 2}{1\over -\Lambda -\pslash} \ .
\ee
$$
The expression \eqCOMPONENT\ of the component fermions
$\psi _{0,\pm \Lambda }$ in terms of the original fermion $\psi $
now explicitly reads
$$
\eqalign{
&\psi _0 = \Lambda ^{-2}(\dirac\dirac+\Lambda ^2)\psi  \cr
&\psi _{\pm \Lambda } = {1\over 2\Lambda ^2}(-\dirac\dirac\pm
i\dirac\Lambda )\psi  \ .\cr
}\eqn\eqARELATION
$$

\subsection{B picture}

In some cases, it is more convenient to choose another form for the
free kinetic term ${\cal L}_{\rm free}$. Indeed when $\Sigma (x)$
develops a
nonvanishing vacuum expectation value (VEV)
$\bra0 \Sigma(x) \ket0 = m {\bf1}_{\nf}\not= 0$,
the fermion acquires a mass term $-m\bar\psi \psi $. In such a case,
we can take the following lagrangian as the free kinetic term by
including the mass term:
$$
{\cal L}^{(m)}_{\rm free} =
{\bar \psi}i\dirac(1+{{\dirac \dirac}\over{\Lambda^2}})\psi
-m{\bar \psi}\psi \ .
\eqn\eqFREEMZ
$$
Then, for this choice, we have
$$
f(x)=x(1-x^2/\Lambda ^2)-m \equiv  -\Lambda ^{-2}(x-m_0)(x-m_1)(x-m_2)
\ ,
\eqn\eqMASSIVEFX
$$
and the kinetic term \eqFREEMZ\ is rewritten into
$$
{\cal L}^{(m)}_{\rm free} =
\eta _0^{-1}\bar\psi _0(i\dirac-m_0)\psi _0 +
\eta _1^{-1}\bar\psi _1(i\dirac-m_1)\psi _1 +
\eta _2^{-1}\bar\psi _2(i\dirac-m_2)\psi _2 \ .
\eqn\eqFREEM
$$
We refer to this as `B picture', which corresponds to the following
decomposition of the fermion propagator:
$$
{1\over m-\pslash(1-{p^2\over \Lambda ^2})} =
\eta _0{1\over m_0-\pslash} +\eta _1{1\over m_1-\pslash}
 +\eta _2{1\over m_2-\pslash} \ .
\eqn\eqBDECOMPOSITION
$$
An inconvenience for analytic treatment in this choice is
that we have no simple explicit expressions for
the three masses $m_j$ (and hence the weights
$\eta _j=[f'(m_j)]^{-1}$ also); they are determined by
Eq.\eqMASSIVEFX\ and their explicit expressions can only be given by
a complicated Cardano's formula.
However, for the purpose to do practical calculations,
we can make use of computer. Then such complication is of no problem
and the decomposition \eqFREEM\ reducing the higher derivative system
into the usual fermion system provides us with very efficient tool to
calculate various quantities, as we shall see explicitly in Sect.3.

We assume henceforth that we are labeling the three masses
$(m_0, m_1, m_2)$ in order such that they approach to
$(0, +\Lambda , -\Lambda )$,
respectively, as the mass parameter $m$ goes to zero. So $\psi _0$
with mass $m_0$ is the physical fermion with positive metric and
the other $\psi _{1,2}$ are unphysical fermions with negative metric.
We should note the fact that possible value
of the physical fermion mass $m_0$ is bounded from above by
$$
m_0 \leq  {\Lambda \over \sqrt 3} = 0.57735\cdots \Lambda  ,
\eqn\eqBOUND
$$
although the mass parameter $m$ can be arbitrarily large
in principle. This is because the root $m_0$ becomes complex
beyond this limit.
Indeed, the cubic polynomial $f(x)$ in \eqMASSIVEFX\
has two stationary points at
$\pm \Lambda /\sqrt 3$ and the root $m_0$, if being real,
has to lie in between them. This bound \eqBOUND\ is not a defect of
the present definition of the NJL model but is of physical
significance. The NJL model is defined with an ultraviolet
cutoff in any case and can only describe physics below it. Therefore
the physical fermion mass $m_0$ generated by the NJL model dynamics
can be at most of the same order as the cutoff and otherwise becomes
unreliable.

The expression \eqCOMPONENT\ of the component fermions
$\psi _j$ in terms of the original one $\psi $ now reads
$$
\myvector{\psi _0}
         {\psi _1}
         {\psi _2}
        = -\Lambda ^{-2} \mymatrix
         {\eta _0} {\eta _0m_0} {-\eta _0\Lambda ^2{m\over m_0}}
         {\eta _1} {\eta _1m_1} {-\eta _1\Lambda ^2{m\over m_1}}
         {\eta _2} {\eta _2m_2} {-\eta _2\Lambda ^2{m\over m_2}}
        \myvector{-\square \psi }
                 {i\dirac\psi }
                 {\psi }
\ , \eqn\eqBRELATION
$$
where we have used the following relations for the three masses
$m_j$ which directly follow from the defining equation \eqMASSIVEFX:
$$
m_0+m_1+m_2=0, \qquad m_0m_1+m_1m_2+m_2m_0=-\Lambda ^2,
\qquad m_0m_1m_2= -m\Lambda ^2 \ .
\eqn\eqMASSRELI
$$
For later convenience, we here cite some formulas for the masses $m_j$
and the weights $\eta _j$. Using the relations \eqMASSRELI\ and
$m_j^3= \Lambda ^2(m_j-m)$ for $\forall j$, one can easily derive
$$
\eqalign{
&\sum _{j=0}^2 m_j^2 = 2\Lambda ^2, \qquad
\sum _{j=0}^2 m_j^3 = -3m\Lambda ^2, \qquad
\sum _{j=0}^2 m_j^4 = 2\Lambda ^4, \cr
&\sum _{i>j} m_i^2m_j^2 = \Lambda ^4, \qquad
{m\over m_j}= 1-{m_j^2\over \Lambda ^2}, \qquad
\hbox{\etc.} \cr
}\eqn\eqMASSRELII
$$
The defining equation of the weights $\eta _j$ together with
these mass relations lead to the following relations:
$$
\eqalign{
& \eta _j = \Big({m\over m_j}-2{m_j^2\over \Lambda ^2}\Big)^{-1}
= \Big(3{m\over m_j}-2\Big)^{-1}, \qquad
\sum _{j=0}^2 \eta _j = 0, \quad
\sum _{j=0}^2 \eta _jm_j = 0, \cr
&\sum _{j=0}^2 \eta _jm_j^2 = -\Lambda ^2, \quad
\sum _{j=0}^2 \eta _jm_j^3 = 0, \quad
\sum _{j=0}^2 {\eta _j\over m_j} = {1\over m}, \quad
\sum _{j=0}^2 {\eta _j\over m_j^2} = {1\over m^2}, \quad
\ \hbox{\etc.} \cr
}\eqn\eqETAREL
$$

\section{Current Operators}

We have emphasized that our definition of NJL model respects the
chiral symmetry. It is indeed clear that our Lagrangian
\eqLAGRANGIAN\ is invariant under the chiral
$U(\nf)_{\rm R}\times U(\nf)_{\rm L}$ transformation. Let us derive
current operators corresponding to this symmetry since they
take different forms from the usual ones because of the presence of
higher derivatives.

To do this, the simplest way is to use the A picture. Then the
Lagrangian is given by \eqFREE\ (plus interaction term containing
no derivatives) to which the usual Noether procedure applies.
In view of the expression \eqARELATION\ for the component fields
$\psi _{0,\pm \Lambda }$ in terms of the original $\psi $,
we see that the vector transformation
$\psi \rightarrow \psi '=e^{i\theta }\psi $
$(\theta \equiv \sum _a\theta ^a\lambda ^a)$ on the original fermion
field $\psi $ is realized on the component fields $\psi _{0,\pm \Lambda
}$
in the same form,  $\psi _{0,\pm \Lambda }\rightarrow \psi _{0,\pm
\Lambda }'=e^{i\theta }\psi _{0,\pm \Lambda }$,
while the chiral transformation
$\psi \rightarrow \psi '=e^{i\gamma _5\theta }\psi $ is realized on the
component fields
a bit differently as follows:
$$
\eqalign{
{\psi'}_0&=\chiral \psi_0\cr
{\psi'}_\Lambda+{\psi'}_{-\Lambda}
&=\chiral(\psi_\Lambda+\psi_{-\Lambda})\cr
{\psi'}_\Lambda-{\psi'}_{-\Lambda}
&=\-chiral(\psi_\Lambda-\psi_{-\Lambda}) \ .\cr
}\ee
$$
Having found the transformation law of the component fields
$\psi _{0,\pm \Lambda }$, we can apply the usual Noether procedure and
obtain the corresponding currents.
The vector current is found to be
$$
j_a^\mu=\bar\psi _0 \lambda ^a\gamma ^\mu \psi _0
-2\bar\psi _\Lambda \lambda ^a \gamma ^\mu \psi _\Lambda
-2\bar\psi _{-\Lambda }\lambda ^a \gamma ^\mu \psi _{-\Lambda } \ ,
\eqn\eqVCURRENT
$$
and the axial current is given by
$$
j_{5a}^\mu=\bar\psi _0 \lambda ^a\gamma ^\mu \gamma _5\psi _0
-2\bar\psi _\Lambda \lambda ^a \gamma ^\mu \gamma _5\psi _{-\Lambda }
-2\bar\psi _{-\Lambda }\lambda ^a \gamma ^\mu \gamma _5\psi _\Lambda  \
{}.
\eqn\eqACURRENT
$$
We can rewrite these current expressions in terms of the original
fermion field $\psi $  by using relations \eqARELATION\
and find
$$
\eqalign{
j_a^\mu  &= \bar\psi \lambda ^a\gamma ^\mu \psi
+ {1\over \Lambda ^2}\big[\square \bar\psi \cdot \lambda ^a\gamma ^\mu
\psi
- \bar\psi \mathop{\dirac}^{\leftarrow }\lambda ^a \gamma ^\mu
\dirac\psi
+ \bar\psi \lambda ^a \gamma ^\mu  \square \psi _{-\Lambda } \big]\ ,
\cr
j_{5a}^\mu  &= \bar\psi \lambda ^a\gamma ^\mu \gamma _5\psi
+ {1\over \Lambda ^2} \big[\square \bar\psi \cdot \lambda ^a\gamma ^\mu
\gamma _5\psi
+ \bar\psi \mathop{\dirac}^{\leftarrow }\lambda ^a \gamma ^\mu \gamma
_5\dirac\psi
+ \bar\psi \lambda ^a \gamma ^\mu \gamma _5 \square \psi _{-\Lambda }
\big]\ . \cr
}\eqn\eqPSICURRENTS
$$

Although not being well-known, there is in fact a direct and general
procedure for deriving the Noether current (without decomposing
$\psi $) for generic higher derivative systems, which
we show in the Appendix.
We can see that the Noether current obtained by the direct procedure
coincides with
the above \eqPSICURRENTS\ up to so-called ambiguity term of the form
$\partial _\nu f^{\mu \nu }$ with an antisymmetric tensor
$f^{\mu \nu }$.

In some cases we need current expression written in terms of the
component fields $\psi _j$ in B picture. This can be obtained if we
invert the relation \eqBRELATION\ between $\psi _j$ and
$( \psi , \dirac\psi , \square \psi  )$:
$$
\myvector{-\square \psi }
         {i\dirac\psi }
         {\psi }
        = \mymatrix
         {m_0^2} {m_1^2} {m_2^2}
         {m_0}   {m_1}   {m_2}
         {1}     {1}     {1}
        \myvector{\psi _0}
                 {\psi _1}
                 {\psi _2}
\ . \eqn\eqBRELATIONINV
$$
Substituting this into \eqPSICURRENTS, we find
the currents in the B picture:
$$
\eqalign{
j_a^\mu  &= \sum _{j=0}^2 \eta _j^{-1}\bar\psi _j\lambda ^a\gamma ^\mu
\psi _j
\ , \cr
j_{5a}^\mu  &=
\sum _{j=0}^2 \eta _j^{-1}\bar\psi _j\lambda ^a\gamma ^\mu \gamma
_5\psi _j +
\sum _{i,j=0}^2 {2m_im_j\over \Lambda ^2}\bar\psi _i\lambda ^a\gamma
^\mu \gamma _5\psi _j
\ .\cr
}\eqn\eqBCURRENTS
$$

\section{Coupling to External Gauge Fields}

It is easy to couple external gauge fields to the system by gauging
the chiral $U(\nf)_{\rm R}\times U(\nf)_{\rm L}$ symmetry.
This is achieved simply by replacing the kinetic term by the
covariant derivative one:
$$
\eqalign{
&\bar\psi  i\dirac \big( 1+{\dirac \dirac\over \Lambda ^2} \big)\psi
\cr &\ \longrightarrow \
\bar\psi  i\Dslash \big( 1+{\Dslash \Dslash\over \Lambda ^2} \big)\psi
 = \bar\psi  i(\dirac-i\Vslash-i\Aslash\gamma _5)
\Big( 1+{(\dirac-i\Vslash+i\Aslash\gamma _5)
(\dirac-i\Vslash-i\Aslash\gamma _5)\over \Lambda ^2} \Big)\psi \ ,
\cr}\ee
$$
where $\calV_\mu =\sum _a \calV_\mu ^a\lambda ^a$ and
$\calA_\mu =\sum _a \calA_\mu ^a\lambda ^a$
are the vector and axial-vector gauge fields of
the chiral $U(\nf)_{\rm R}\times U(\nf)_{\rm L}$ group. If we gauge a
part of the group, the gauge fields should also be understood so;
for instance, if we couple only the photon $A_\mu $ to the system,
then $\calV_\mu =eQA_\mu $ and $\calA_\mu =0$ with $Q$ being the
charge quantum number matrix of the fermion.
Note that the covariant derivative $D_\mu $  of course
depends on the operand and
$D_\mu \psi =(\partial _\mu -i\calV_\mu -i\calA_\mu \gamma _5)\psi $
while
$D_\mu \Dslash \psi =(\partial _\mu -i\calV_\mu +i\calA_\mu \gamma
_5)\Dslash\psi $.

Here two remarks are in order. First we note that the above
covariant kinetic term contains
the terms linear in the gauge fields in the form,
$$
\calV_\mu ^aj_a^\mu  + \calA_\mu ^aj_{5a}^\mu
+ (\hbox{total derivatives}),
$$
and the currents $j_a^\mu $ and $j_{5a}^\mu $ just coincide with the
above ones given in \eqPSICURRENTS.  Second, in the above
covariantization we started with the expression
$\bar\psi  i\dirac \big( 1+(\dirac \dirac/\Lambda ^2) \big)\psi $.
But if we started with an equivalent one
$\bar\psi  i\dirac \big( 1+(\square /\Lambda ^2) \big)\psi $,
we would have obtained
$\bar\psi  i\Dslash \big( 1+(D^2/\Lambda ^2) \big)\psi $,
which is different from the above one
$\bar\psi  i\Dslash \big( 1+(\Dslash\Dslash/\Lambda ^2)\big)\psi $.
The difference is, however, seen to be a non-minimal interaction
term (Pauli term) of the form like
$F_{\mu \nu }\bar\psi \gamma ^\mu \gamma ^\nu \psi $
with a field strength $F_{\mu \nu }$,
and hence the currents defined by the linear terms in the gauge fields
again coincides with the above ones up to ambiguity
terms of the form $\partial _\nu f^{\mu \nu }$.

We have seen that our definition of the NJL model
by higher derivative is made consistent with the gauge symmetry also.
However, we should note the fact that the vertex functions of the
external gauge fields alone are not yet well-regularized by our higher
derivative fermion kinetic term.  Consider generally
a fermion one-loop diagram which contains $n$ vertices of
`mesons' $\Sigma $, $R_\mu $ and $L_\mu $
and $m$ vertices of external gauge fields $\calV_\mu $ and $\calA_\mu
$.
The fermion propagator behaves as $\sim k^{-3}$ at high loop
momentum $k$ and the vertex factor for `mesons' contains no momentum.
But the point is that the vertex of the external gauge fields
contains second powers of fermion momenta as the currents
\eqPSICURRENTS\ show. Therefore the diagram
has superficial degree of divergence
$$
\omega =4-3(n+m)+2m = 4-3n-m .
$$
This is non-negative when $(n=1, m=0,1)$ and $(n=0, m=1,2,3,4)$
aside from an irrelevant vacuum graph case ($n=m=0$). The linear or
logarithmic divergences for the former cases $(n=1,m=0,1)$
are in fact absent because
of Lorentz, chiral and gauge invariances.
Indeed, first, the $(n=1, m=0)$ case corresponds to `meson' tadpoles
(1-point functions),
and $\Sigma $-tadpole vanishes by the chiral symmetry while
vector $R_\mu $ and $L_\mu $ tadpoles vanish by Lorentz invariance.
Second, the $(n=1, m=1)$ case corresponds to `meson'-gauge transition
2-point vertex functions;
when the meson is $\Sigma $, the 2-point vertex $\Sigma $-$\calV_\mu $
or
$\Sigma $-$\calA_\mu $ vanishes by chiral symmetry,
and when the meson is $R_\nu $ or
$L_\nu $, the 2-point vertex like $(R+L)_\nu $-$\calV_\mu $ should be
proportional to $(g_{\mu \nu }p^2-p_\mu p_\nu )$ by gauge invariance.
Therefore they, being superficially logarithmically divergent,
become convergent actually. The divergences thus
occur only when $(n=0, m=1,2,3,4)$, \ie,
for the {\it diagrams consisting of external gauge fields alone}.
For those diagrams the present higher derivative kinetic term  does
not improve the divergence situation at all; however higher the
fermion kinetic term derivatives is chosen, the gauge boson vertex
also gets to contain higher derivatives accordingly. This is a
well-known fact in the higher derivative
regularization\rlap.\refmark{\Slavnov}  We can
however regularize those divergence using dimensional or Pauli-Villars
or any other suitable regularization\rlap.\foot{
Note that this is a mere regularization which is to be removed
eventually by the usual renormalization.}
For definiteness, we adopt dimensional regularization
in this paper.

\chapter{Example Calculations}

\section{Effective Action and Potential}

As we have seen in the above, the naive Feynman rule is correct
for this type of higher derivative system. This also implies that
the usual Feynman path integral expression for the Green function
generating functional is valid for our system with lagrangian ${\cal
L}$
given by \eqLAGRANGIAN. Therefore the effective action $\Gamma $ for
the
`meson' fields $\Sigma , \Sigma ^\dagger , R_\mu $ and $L_\mu $ is
given in the
leading order in $1/\Nc$ simply by integrating over the
fermion field $\psi $ (and $\bar\psi $):
$$
\eqalign{
\Gamma  = \int d^4x &\ [\tree] \cr
&\qquad +{{\Nc}\over i}
\Tr\Ln[i\dirac(1+{{\dirac
\dirac}\over{\Lambda^2}})-\calM] \ .\cr
} \eqn\eqEFFACTION
$$
[Here we have omitted the last term in \eqLAGRANGIAN,
$(\Lambda ^2/4)\left( g_\VZ^{-2}- g_\V^{-2}\right)
\big(\left[\tr(R_\mu \lambda ^0)\right]^2 + \left[\tr(L_\mu \lambda
^0)\right]^2 \big) $
simply for brevity of writing. Namely we may understand that we
consider only the $g_\V^{}=g_\VZ^{}$ case henceforth, or otherwise,
it should be understood that this vector singlet term is always
accompanying the vector term $(\Lambda ^2/2)\tr(R^\mu R_\mu +L^\mu
L_\mu )$.]\
Precisely speaking, the second $\Tr\Ln$ term standing for the
fermion one-loop diagrams is not made convergent yet by
our third order derivative propagator. But if we expand it with respect
to $\calM$, the ultraviolet divergence appears only in the first two
terms; the zero-th and first order terms in $\calM$. Since the first
order term vanishes by chiral symmetry and Lorentz invariance as we
explained above, the
divergence in fact appears only in the zero-th order term which
is a field independent constant we can throw away freely.

The effective potential $V(\Sigma )$, whose stationary point
determines the VEV of
$\Sigma $, is easily found from \eqEFFACTION\ to be given by
$$
\eqalign{
V(\Sigma )
&= {\Lambda ^2\over 2g_\S^2}\tr(\Sigma ^\dagger \Sigma )
 - \Nc \int {d^4p\over i(2\pi )^4}
  \tr\ln\big[\Sigma \calPR+\Sigma ^\dagger \calPL
        -\pslash\big(1-{\pslash\pslash\over \Lambda ^2}\big) \big] \cr
&= {\Lambda ^2\over 2g_\S^2}\tr(\Sigma ^\dagger \Sigma )
 - 2\Nc \int {d^4p\over i(2\pi )^4} \tr\ln\big[\Sigma ^\dagger \Sigma
        -p^2\big(1-{p^2\over \Lambda ^2}\big)^2 \big] \ , \cr
}\ee
$$
where tr in the second integral expression denotes the trace
only over the flavor space.
The scalar field $\Sigma $ is generally expanded into
$$
\Sigma  = \sigma  + i\pi \ , \ \ \ \
 \qquad \sigma  = \sum _{a=0}^{\nf^2-1}\sigma ^a {\lambda ^a\over 2}\ ,
 \qquad \pi  = \sum _{a=0}^{\nf^2-1}\pi ^a {\lambda ^a\over 2} \ ,
\ee
$$
and we see that the VEV of $\Sigma $ is always brought into a real and
diagonal matrix form by the chiral $U(\nf)_{\rm R}\times U(\nf)_{\rm L}$
rotation and further that the diagonal matrix is in fact proportional
to unit matrix since each of the diagonal values is separately
determined by minimizing exactly the same form of potential. Therefore,
without of loss of generality, we can substitute
$$
\VEV{\Sigma } = \sigma ^0 \lambda ^0 \equiv  m {\bf1}_{\nf}
\ee
$$
into the effective potential, where $m$ is a mass
parameter which will yield a fermion mass term $-m\bar\psi \psi $.
Then we find the effective potential to be
$$
\eqalign{
{1\over \nf}V\big(\Sigma = m{\bf1}_{\nf}\big)
&= {\Lambda ^2\over 2g_\S^2}m^2
 - 2\Nc \int {d^4p\over i(2\pi )^4} \ln\big[
 m^2-p^2\big(1-{p^2\over \Lambda ^2}\big)^2 \big] \ . \cr
}\eqn\eqEFFECTIVEPOT
$$
The stationary condition of the potential leads to
the following gap equation which determines the non-zero mass
value $m$ corresponding to spontaneous chiral symmetry breaking:
$$
{\Lambda ^2\over 4g_\S^2\Nc}- \int {d^4p\over i(2\pi )^4}
{1\over m^2-p^2\big(1-{p^2\over \Lambda ^2}\big)^2} = 0 \ .
\eqn\eqGAPEQ
$$

Up to here the story is formally the same as in the usual NJL
model case and the difference is only in the higher derivative
term in the fermion propagator.
As noted in the previous section, the higher derivative propagator
can be reduced to the sum of ordinary fermion propagators and then
we can evaluate the quantities in the same manner as
usual. For instance, consider the effective potential
in \eqEFFECTIVEPOT. If we use
Eq.\eqMASSIVEFX\ with $x=\pm \pslash$ substituted,
we can evaluate it as
$$
{1\over \nf}V\big(\Sigma = m{\bf1}_{\nf}\big)
= {\Lambda ^2\over 2g_\S^2}m^2  + \Nc \sum _{j=0}^2F_0(m_j)\ ,
\eqn\eqPOTPOT
$$
where $m_j$ are the three masses of the component fermion $\psi _j$,
determined by Eq.\eqMASSIVEFX, and the
function $F_0(m_j)$ is the usual fermion vacuum energy given by
$$
\eqalign{
F_0(m_j) &=  - \int {d^np\over i(2\pi )^n} \tr\ln(m_j-\pslash)
 =  -2\int {d^np\over i(2\pi )^n} \ln(m_j^2-p^2) \cr
 &=  - {1\over 16\pi ^2}m_j^4\left(\VL + \ln{m_j^2\over \Lambda
^2}-{3\over 2}\right)
                    + O(\epsilon )\ , \cr
}\ee$$
with $\VL$ standing for the `divergent' part:
$$
\eqalign{
&\VL \equiv  \ln{\Lambda^2}-{1\over{\bar \epsilon }} , \cr
&{1\over{\bar\epsilon }}\equiv {1\over\epsilon }-\gamma +\ln(4\pi),
\qquad
\epsilon \equiv  {4-n\over2}, \qquad \gamma :\ \hbox{Euler constant} \
. \cr
}\eqn\eqDIVFAC
$$
Here note that we have used dimensional regularization to evaluate
the contributions from the component fermions. This is always
necessary since those contributions are separately
ultraviolet-divergent, although the sum is generally convergent.
In this case of effective potential, however,
the sum still contains an $m$-independent divergence as mentioned
above. In fact, the sum $\sum _{j=0}^2 F(m_j)$ contains the divergent
part $-(\VL/16\pi ^2)\sum _jm_j^4$ and it is seen to be $m$-independent
constant because of an equality $\sum _jm_j^4= 2\Lambda ^4$ in
\eqMASSRELII.
Discarding the $m$-independent divergence and constants, we find the
potential \eqPOTPOT\ to yield
$$
{1\over \nf}V\big(\Sigma = m{\bf1}_{\nf}\big)
= {\Lambda ^2\over 2g_\S^2}m^2  -
{\Nc\over 16\pi ^2} \sum _{j=0}^2 m_j^4\ln{m_j^2\over \Lambda ^2} \ .
\eqn\eqPOTPOTPOT
$$
The constant is adjusted so that this becomes zero as $m\rightarrow 0$
in which $(m_0, m_1, m_2)$  $\rightarrow $  $(0, +\Lambda , -\Lambda
)$.
The gap equation \eqGAPEQ\ can also be rewritten as follows
if we use an equality
$\big[m^2-p^2\big(1-(p^2/\Lambda ^2)\big)^2\big]^{-1}$
$ = \sum _j \eta _j(m_j/m)[m_j^2-p^2]^{-1}$
which follows from taking trace of both sides of Eq.\eqBDECOMPOSITION:
$$
{\Lambda ^2\over 4g_\S^2\Nc} = {1\over 16\pi ^2}\sum _{j=0}^2
\eta _j\,{m_j^3\over m}\,\ln{m_j^2\over \Lambda ^2} \ .
\ee
$$

\section{Two-point Functions}

The same technique applies also to other quantities. For instance,
the $n$-point Green functions on the spontaneously broken vacuum is
obtained by expanding the effective action \eqEFFACTION\ around
$\calM=m$. So the fermion propagator is given by
$i[\pslash(1-p^2/\Lambda ^2)-m]^{-1}$ which can be decomposed into the
same component fermion propagators as above. Let us demonstrate this
by calculating the 2-point functions $\Gamma ^{(2)}_{\sigma \ {\rm or}\
\pi }$
of scalar $\sigma $ and
pseudoscalar $\pi $ on this vacuum.
For every flavor components $\sigma ^a$ and $\pi ^a$,
independently of $a=0,1,\cdots ,\nf^2-1$,  they are given by
$$
\eqalign{
\Gamma ^{(2)}_{\left\{{\scriptstyle\sigma  \atop \scriptstyle\pi
}\right\}}(p^2)
 = &-{\Lambda ^2\over 2g_\S^2} \cr
  &-{\Nc\over 2}\int {d^4k\over i(2\pi )^4} \tr
 \Big[ \left\{{1 \atop i\gamma _5}\right\}{1\over m-\reg{\kslash}}
 \left\{{1 \atop i\gamma _5}\right\}{1\over m-\reg{\kslash+\pslash}}
        \Big] \ , \cr
}\eqn\eqTWOPOINT
$$
where $\reg{\kslash}$ is a shorthand notation for
$\kslash\big(1-k^2/\Lambda ^2\big)$.
The second term standing for the fermion one-loop diagram can be
evaluated by using the B picture decomposition \eqBDECOMPOSITION\
for each propagator as follows:
$$
\eqalign{
\hbox{second}& \hbox{ term} \cr
  &= -{\Nc\over 2}\sum _{i,j=0}^2\eta _i\eta _j\int {d^nk\over i(2\pi
)^n} \tr
 \Big[ \left\{{1 \atop i\gamma _5}\right\}{1\over m_i-\kslash}
 \left\{{1 \atop i\gamma _5}\right\}{1\over m_j-(\kslash+\pslash)}
\Big]
\cr
 &= {\Nc\over 8\pi ^2}\sum _{i,j=0}^2\eta _i\eta _j
 \Big[ \big(m_i^2+m_j^2\pm m_im_j-{p^2\over 2}\big)\VL
 -\big({m_i^2+m_j^2\over 2}-{p^2\over 6}\big) \cr
 & \hskip7cm
 + F_{\pm }(p^2; m_i, m_j) \Big] \ , \cr
}\eqn\eqSECONDTERM
$$
with
$$
\eqalign{
F_{\pm }(p^2; m_i, m_j)
&\equiv  \int _0^1dx \left(2\Mij\pm m_im_j-3x(1-x)p^2\right)
       \ln{\Mij-x(1-x)p^2\over \Lambda ^2}\ , \cr
\Mij &\equiv  (1-x)m_i^2+xm_j^2 \ . \cr
}\eqn\eqFUNCTIONPM
$$
All the divergent terms proportional to $\VL$ in \eqSECONDTERM\
are seen to vanish if we use the identities
$\sum _{j=0}^2 \eta _j = 0$ and $\sum _{j=0}^2 \eta _jm_j = 0$ in
\eqETAREL\
and therefore we obtain
$$
\Gamma ^{(2)}_{\left\{\!{\scriptstyle\sigma  \atop \scriptstyle\pi
}\!\right\}}(p^2)
=-{\Lambda ^2\over 2g_\S^2}
+ {\Nc\over 8\pi ^2}\sum _{i,j=0}^2\eta _i\eta _j
 F_{\pm }(p^2; m_i, m_j)\ .
\eqn\eqTWOPOINTSP
$$
In quite the same way we can calculate 2-point functions for the
vector and axial vector mesons, $V_\mu \equiv (R_\mu +L_\mu )/2$ and
$A_\mu \equiv (R_\mu -L_\mu )/2$: assuming $g_\V^2=g_\VZ^2$, they are
given independently of the flavor by
$$
\eqalign{
&\Gamma ^{(2)\,\mu \nu }
_{\left\{\!{\scriptstyle V \atop \scriptstyle A}\right\}\!}(p^2)
\cr
&= g^{\mu \nu }{\Lambda ^2\over g_\V^2}
  -{\Nc\over 2}\sum _{i,j=0}^2 \! \eta _i\eta _j
\int \!{d^4k\over i(2\pi )^4} \tr
 \Big[ \left\{\!{\gamma ^\mu  \atop \gamma ^\mu \gamma _5}\!\right\}
 {1\over m_i-{\kslash}}
 \left\{\!{\gamma ^\nu  \atop \gamma ^\nu \gamma _5}\!\right\}
 {1\over m_j-(\kslash+\pslash)} \Big]  \cr
&= g^{\mu \nu }\Big[{\Lambda ^2\over g_\V^2}
  -{\Nc\over 8\pi ^2}\sum _{i,j}\eta _i\eta _j
G_{\left\{\!{\scriptstyle V \atop \scriptstyle
A}\!\right\}}(p^2;m_i,m_j)
\Big] \cr
&\ \qquad \qquad \qquad \ \ \ \
+(g^{\mu \nu }p^2-p^\mu p^\nu ){\Nc\over 8\pi ^2}
  \sum _{i,j}\eta _i\eta _j H(p^2;m_i,m_j)
\ ,\cr
}\eqn\eqTWOPOINTVA
$$
where
$$
\eqalign{
G_{\left\{\!{\scriptstyle V \atop \scriptstyle
A}\!\right\}}(p^2;m_i,m_j)
&\equiv  \int _0^1dx \left(\Mij\mp m_im_j\right)
       \ln{\Mij-x(1-x)p^2\over \Lambda ^2}\ , \cr
H(p^2;m_i,m_j)
&\equiv  \int _0^1dx\, 2x(1-x)\ln{\Mij-x(1-x)p^2\over \Lambda ^2}\ . \cr
}\eqn\eqFUNCTIONVA
$$
Although these Eqs.\eqTWOPOINTSP\ and \eqTWOPOINTVA\
are not so explicit expressions since the roots $m_j$
are implicit functions of $m$ and $\Lambda $,
they give closed expressions for the 2-point functions
which allows straightforward numerical evaluations. (In particular,
the parameter integrals over $x$ in \eqFUNCTIONPM\ and \eqFUNCTIONVA\
can be carried
out analytically and the functions $F_{\pm }$,
$G_{\left\{\!{\scriptstyle V \atop \scriptstyle A}\!\right\}}$ and
$H$ are given explicitly by elementary functions.)

We here note some points on these 2-point functions. Firstly
the pion is indeed massless, $\Gamma ^{(2)}_\pi (p^2\!=\!0)=0$,
as it should be because of chiral symmetry. This can immediately
be seen by putting $p=0$ directly in \eqTWOPOINT\ and comparing it
with the gap equation \eqGAPEQ.
[It is more complicated to see this if we use the expression
\eqTWOPOINTSP, although being possible by the help of a formula
$\sum _j\eta _j/(m_j+m_i) = 1/2m\ (m_i\hbox{-independent})$
following from the relation
$\sum _j\eta _j/(m_j-x) = -\Lambda ^2/\prod _j(m_j-x)$.]\ \
Thus the pion 2-point function behaves as
$\Gamma ^{(2)}_\pi (p^2) = Z_\pi ^{-1}p^2 + O(p^4)$ around $p^2=0$
and the coefficient $Z_\pi ^{-1}$ defines the (inverse of)
wave-function renormalization factor of our $\pi $ field
as $\pi =Z_\pi ^{{1\over 2}}\pi _{\rm r}$. Since
there is no genuine Yukawa vertex correction in the $1/\Nc$
leading order, this wave-function renormalization factor also
determines the Yukawa coupling $\Yukawa$ defined by
${\cal L}_{\rm int}=-\Yukawa\bar\psi i\gamma _5\pi _{\rm r}\psi $;
namely,
$\Yukawa=Z_\pi ^{1\over 2}$.
Evaluating the coefficient of $p^2$ in \eqTWOPOINTSP,
we find
$$
Z_\pi ^{-1}=\Yukawa^{-2}=
{\Nc\over 8\pi ^2}\sum _{i,j=0}^2\eta _i\eta _j
{\partial F_{-}\over \partial p^2}(0;m_i,m_j) \ ,
\eqn\eqZPI
$$
$$
\eqalign{
{\partial F_{-}\over \partial p^2}(0;m_i,m_j) &=
\bigg[{-m_i^3(m_i+2m_j)\ln{m_i^2\over \Lambda ^2}
    \over  2(m_i+m_j)^3(m_i-m_j)}
 + \big(i\leftrightarrow j\big)\bigg]
 +{m_im_j\over 2(m_i+m_j)^2}+{1\over 12} \cr
&\quad \bigg( \ \ \mathop{\longrightarrow }_{m_i\rightarrow m_j}\
-{1\over 2}\ln{m_i^2\over \Lambda ^2} - {1\over 6} \ \ \bigg) \ .\cr
}\ee
$$

Secondly, an interesting point in our definition of NJL model
is that the scalar $\sigma $ meson develops a pole {\it below} the
two fermion threshold $2m_0$. Namely, the famous NJL mass relation
$m_\sigma =2m_0$ no longer holds here. We can find the $\sigma $ meson
mass
numerically as a zero of $\Gamma _\sigma ^{(2)}$ in \eqTWOPOINTSP\ and
show
the result in Fig.1.
Note that the NJL mass relation holds only in the limit
$m_0/\Lambda  \rightarrow  0$.
[The approach to the NJL mass relation point is rather singular
and is in fact difficult to be traced numerically.
But analytic estimation gives the behavior
$$
{m_\sigma ^2\over m_0^2} \simeq  4 - {6\over \ln(\Lambda ^2/ m_0^2)}
$$
to a good accuracy in the narrow region
$(m_0/\Lambda )\lsim e^{-5}=0.0067\cdots $. ]
This phenomenon that the NJL mass relation is realized in the limit
$m_0/\Lambda  \rightarrow  0$ is in accord with the other authors'
observation that
the NJL mass relation is realized as an infrared fixed point of
the renormalization group flow\rlap.\refmark{\Maekawa}
In any case, we suspect from this that the NJL mass relation is not so
universal relation proper to the NJL model.
\figinsert{1}{masses.eps}{12cm}
    {Meson mass squares $m_b^2$ vs fermion mass $m_0$.
     $\sigma, \ V$ and $A$ denote scalar, vector and axial-vector
     meson boundstates, respectively. ($m_0/\Lambda$ is bounded by
     $1/\sqrt{3}$.)}

Finally, as for the vector and axial-vector meson 2-point functions
\eqTWOPOINTVA,
we note that the one-loop contribution parts
do not have gauge-invariant form
$\propto (g_{\mu \nu }p^2-p_\mu p_\nu )$
even in the limit $m\rightarrow 0$ contrary to the true gauge fields.
This is of course because the apparent `gauge-invariance' for those
mesons is violated in the higher derivative terms in the present model.
We have also plotted in Fig.1 the masses of the vector and
axial-vector mesons determined as zeros of $\Gamma ^{(2)}_{V,A}$
in \eqTWOPOINTVA\ for the case $g_\S^2=g_\V^2$ corresponding to
QCD analogue model. It may be of some interest to note that, if we
take $\Lambda= 1$GeV, 900MeV, 800MeV and $f_\pi=93$MeV as inputs,
the present NJL model gives the following values for
the fermion mass $m_0$ and the fermion pair VEV:
$$
\eqalign{
\Lambda= 1\,{\rm GeV} &: \quad m_0 = 250\,{\rm MeV},
\quad\big( -\VEV{\bar\psi\psi}_{\rm 1GeV}
\big)^{1/3} = 253\,{\rm MeV} \cr
\Lambda= 900\,{\rm MeV} &: \quad m_0 = 273\,{\rm MeV},
\quad\big( -\VEV{\bar\psi\psi}_{\rm 1GeV}
\big)^{1/3} = 245\,{\rm MeV} \cr
\Lambda= 800\,{\rm MeV} &:  \quad m_0 = 315\,{\rm MeV},
\quad\big( -\VEV{\bar\psi\psi}_{\rm 1GeV}
\big)^{1/3} = 237\,{\rm MeV} \ ,\cr
}\ee
$$
where Eq.(4.7) below for $f_\pi$ and Eqs.\eqZPI, \eqGAPEQ\ and
\eqEQMOT\ are used.
Eq.\eqEQMOT\ gives $-\VEV{\bar\psi\psi}=(\Lambda^2/g_\S^2)m$, which we
regard as the fermion pair VEV renormalized at the cutoff scale
$\Lambda$. The fermion pair VEV's cited here are renormalized ones
at $\mu=1\,$GeV which we calculated using the following formula
although the renormalization effects are small:
$$
-\VEV{\bar\psi\psi}_\mu
= \bigg({\ln(\mu^2/\Lambda_{\rm QCD}^2)
\over \ln(\Lambda^2/\Lambda_{\rm QCD}^2)} \bigg) ^{4/9}
{\Lambda^2\over g_\S^2}m
\ee
$$
with $\Lambda_{\rm QCD}\simeq 500\,$MeV, where 4/9 is the anomalous
dimension of $\bar\psi\psi$  in three flavored $SU(3)$ QCD.
The empirical value for
$\big( -\VEV{\bar\psi\psi}_{\rm 1GeV}\big)^{1/3}$ to be compared
with these is $225\pm 25\,$MeV.

\chapter{Chiral Symmetry and Low Energy Theorem}

\section{Decay Constant}

The decay constant $f_\pi $ is defined by
$$
\bra{0}\,{1\over 2}j_{5a}^\mu (x)\ |\pi ^b({\bf p})\rangle  =
-if_\pi p^\mu e^{-ipx}\delta _a^b \ ,
\ee
$$
where the factor $1/2$ is put since $(1/2)j_{5a}^\mu $ coincides with
the usual axial current which is defined with $\lambda ^a/2$ inserted
as
the flavor matrix. This matrix element can be evaluated
by the Feynman diagram drawn in Fig.2.
If we use the original field $\psi $, the axial current $j_{5a}^\mu $
is given by Eq.\eqPSICURRENTS\ and therefore the diagram is
read to yield
$$
-if_\pi p^\mu
= i{\Yukawa\Nc\over 2}\int {d^4k\over i(2\pi )^4}
\tr\left[ {1\over m-\reg{\kslash} } \AVF
 {1\over m-\reg{\kslash+\pslash} } \gamma _5 \right]\ ,
\eqn\eqFPI
$$
where $\AVF$ is the axial
current vertex factor given by
$$
\AVF = \gamma ^\mu \gamma _5\big[ 1-{1\over \Lambda
^2}\big(k^2+(k+p)^2\big) \big]
        +{1\over \Lambda ^2}\kslash\gamma ^\mu \gamma
_5(\kslash+\pslash) \ .
\ee
$$
\figinsert{2}{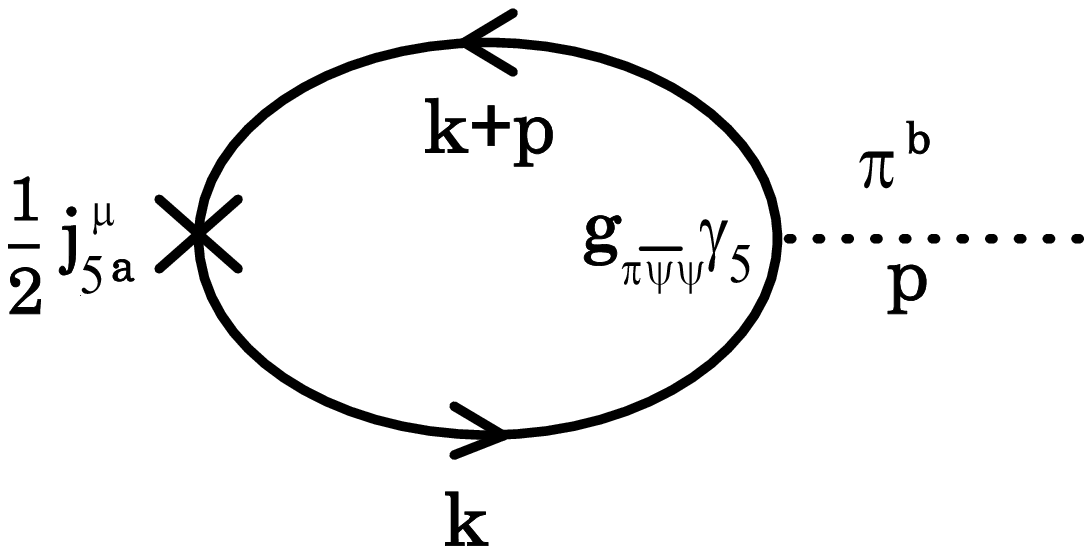}{8cm}
    {Feynman diagram contributing to $f_\pi$.}

Note that the RHS of Eq.\eqFPI\ should be evaluated at the mass
shell $p^2=0$ in order to give the decay constant $f_\pi $. But the
RHS itself, corresponding to the diagram in Fig.2,
{\it is} defined even off the mass shell and hence
the Eq.\eqFPI\ may be
regarded as defining an `off-shell' decay constant $f_\pi (p^2)$
which coincides with the true $f_\pi $ at $p^2=0$.
With this understanding, let $p^\mu $ be off the mass shell for a
while and multiply both sides of Eq.\eqFPI\ by $ip_\mu $. Then
we can obtain
$$
\eqalign{
f_\pi (p^2)\,p^2
= {\Yukawa\Nc\over 2}\int {d^4k\over i(2\pi )^4}  \Bigg\{
&-\tr\left[ {1\over m-\reg{\kslash+\pslash}} + {1\over m-\reg{\kslash}}
\right]
 \cr
&\qquad\ \  +2m \tr\left[{1\over m-\reg{\kslash}} \gamma _5
 {1\over m-\reg{\kslash+\pslash}} \gamma _5 \right]\Bigg\} \ ,
\cr}\eqn\eqFPITWO
$$
by using an algebraic identity
$$
\eqalign{
p_\mu \AVF &= \reg{\kslash+\pslash}\gamma _5 +\gamma _5\reg{\kslash}
\cr
  & = \big(m-\reg{\kslash}\big)\gamma _5
   + \gamma _5\big(m-\reg{\kslash+\pslash}\big) -2m\gamma _5
\ . \cr
}\eqn\eqAXIALWT
$$
Note that this identity \eqAXIALWT\
is just a higher derivative case version of the usual tree level
Ward-Takahashi identity for the axial vector current.
The two terms in the first trace in \eqFPITWO\ become equal to
each other by shifting the loop momentum (which is allowed now)
and they give
$-2m(\Lambda ^2/g_\S^2\Nc)\times (\Yukawa\Nc/2)$
$=$ $-m\Yukawa\Lambda ^2/g_\S^2$ owing to the gap equation \eqGAPEQ.
So we find
$$
f_\pi (p^2)\,p^2
= 2m\Yukawa \Bigg\{ -{\Lambda ^2\over 2g_\S^2} +{\Nc\over 2}
 \int {d^4k\over i(2\pi )^4} \tr\left[{1\over m-\reg{\kslash}} \gamma
_5
 {1\over m-\reg{\kslash+\pslash}} \gamma _5 \right]\Bigg\} \ .
\ee
$$
But the quantity in the curly bracket here is just the same as
the 2-point function $\Gamma _\pi ^{(2)}(p^2)$ in Eq.
\eqTWOPOINT\ of pseudoscalar
$\pi $, which we know behaves as $Z_\pi ^{-1}p^2 + O(p^4)$ around
$p^2=0$. Therefore, taking also account of the relation
$\Yukawa=Z_\pi ^{{1\over 2}}$, we find
$$
f_\pi (p^2\!=\!0)=f_\pi = 2m\Yukawa Z_\pi ^{-1}
= 2m Z_\pi ^{-{1\over 2}} = {2m\over \Yukawa} \ .
\eqn\eqRELFPIYUKAWA
$$

Although we have derived this relation \eqRELFPIYUKAWA\ by a
rather explicit computation in the leading order in $1/\Nc$,
it is in fact a direct consequence of the chiral symmetry alone.
Indeed, using canonical commutation in A (or B) picture
and expressing $\pi $ and $\sigma $ fields in terms of fermion field,
we can derive the chiral
symmetry transformation law of the NG boson field $\pi ^b$:
$$
\delta (x^0) [ ij_{5a}^{0}(x), \,\pi ^b(0)]
= \delta ^4(x) \tr\left[\lambda ^b
             \big(\lambda ^a\sigma (0)+\sigma (0)\lambda
^a\big)\right]\ .
\ee
$$
Then, by using VEV of this equation and
$\bra{0}\sigma (x)\ket{0}= m{\bf1}_{\nf}$,
we see that
the current conservation $\partial _\mu j_{5a}^\mu (x)=0$
leads to
$$
i\partial _\mu   \bra{0} {\rm T}\,j_{5a}^\mu (x) \pi ^b(0) \ket{0}
= \delta (x^0) \bra{0} [ ij_{5a}^{0}(x), \,\pi ^b(0)] \ket{0}
 = 4m \delta ^4(x) \delta _a^b \ .
\ee
$$
This implies that the 2-point function
$\bra{0} {\rm T}\,j_{5a}^\mu  \pi ^b \ket{0}$
in momentum space contains a massless pole and is given by
$(p^\mu /p^2) 4m\delta _a^b $. But, on the other hand,
such massless pole
comes from the NG boson intermediate state and hence
the pole residue should be $2f_\pi p^\mu \cdot Z_\pi ^{{1\over
2}}\delta _a^b$.
We thus get an equality
$2f_\pi Z_\pi ^{1/2} = 4m$,
the same relation as the Eq.\eqRELFPIYUKAWA.

\section{$\pi ^0 \rightarrow  2\gamma $ Amplitude}

We now turn to calculate the amplitude of $\pi ^0 \rightarrow  2\gamma
$ decay.
The neutral pion $\pi ^0$ corresponds to $Z_\pi ^{{1\over 2}}\pi ^3$ in
the
present notation and the photon $\gamma $ couples to ($e$ times)
the vector current
$j_a^\mu (x)$ with the flavor matrix $\lambda ^a$ replaced by the quark
charge matrix $Q = {\rm diag}(2/3, -1/3, \cdots  )$.
The simplest way for calculating this amplitude is to use the
B picture since the vector current is diagonal with respect
to the index $j$ of the component fermions $\psi _j$ even in the
B picture.
The amplitude is calculated by a triangle diagram and is
given by
$$
\eqalign{
&\Mpgg = -e^2\Yukawa\Nc\tr\big(QQ{\lambda ^3\over 2}\big)
\sum _{j=0}^2\eta _j
T^{\mu \nu }(p,q; m_j) \ ,\cr
&T^{\mu \nu }(p,q; m_j) =
 \int {d^nk\over i(2\pi )^n} \Bigg\{
 \tr\left[{1\over m_j-(\kslash-\qslash)}\gamma ^\nu
 {1\over m_j-\kslash}\gamma ^\mu {1\over
m_j-(\kslash+\pslash)}(-i\gamma _5)\right]
 \cr
 & \hskip8cm + \big[(p,\mu ) \leftrightarrow  (q,\nu )\big] \Bigg\} \ .
\cr
}\ee
$$
The evaluation of the integral of the amplitude
$T^{\mu \nu }(p,q; m_j)$ is well known;\refmark{\Jackiw} it gives
$$
T^{\mu \nu }(p,q; m_j) =
-{1\over 4\pi ^2m_j}\epsilon ^{\mu \nu \alpha \beta }q_\alpha p_\beta
\ ,
\ee
$$
so that we find
$$
\eqalign{
\Mpgg &= {e^2\Yukawa\Nc\over 4\pi ^2}
\tr\big(QQ{\lambda ^3\over 2}\big)
\Big(\sum _{j=0}^2{\eta _j\over m_j}\Big)
\epsilon ^{\mu \nu \alpha \beta }q_\alpha p_\beta  \cr
 &= {e^2\Nc\over 4\pi ^2}\big({\Yukawa \over 2m}\big)
\tr(QQ\lambda ^3)\epsilon ^{\mu \nu \alpha \beta }q_\alpha p_\beta  \ .
\cr
}\ee
$$
Here we have used $\sum _j\eta _j/m_j = 1/m$ in Eq.\eqETAREL.
If $\Yukawa/2m = 1/f_\pi $, this exactly reproduces the well-known
low energy theorem for $\pi ^0\rightarrow 2\gamma $. This is indeed the
case
because of the relation \eqRELFPIYUKAWA\
as we confirmed in the above.

\chapter{Simplified Treatment for the Case $\Lambda \gg m$}

In some applications of NJL model (as in the top condensation
scenario\rlap,\refmark{\MTY} for instance),
the cutoff $\Lambda $ is much larger
than the scale of the mass $m$ and momenta which we discuss.
In such cases we need not keep terms which are suppressed by
$1/\Lambda ^2$ in the effective action. Then we can take $\Lambda
^2\rightarrow \infty $
limit in all the terms not diverging in that limit.
This simplifies the calculations considerably and
the calculated results become very explicit ones containing
no longer the implicit masses $m_j$.

We now show how to evaluate the effective action \eqEFFACTION\
in such cases. Only problem is the fermion one-loop term,
which we expand as follows using the notation
$\reg{i\dirac}\equiv i\dirac(1+{\dirac\dirac/\Lambda ^2})$:
$$
\eqalign{
\Tr\Ln\big[\,\reg{i\dirac}-\calM\big]
=\Tr\Bigg[
 \Ln\big[\,\reg{i\dirac}\big]
 &-{\calM\over  \reg{i\dirac}}
 -{1\over 2}\Big({\calM\over  \reg{i\dirac}}\Big)^2
 -{1\over 3}\Big({\calM\over  \reg{i\dirac}}\Big)^3 \cr
 &-{1\over 4}\Big({\calM\over  \reg{i\dirac}}\Big)^4
 -\left(\sum _{n\geq 5}{1\over n}\Big({\calM\over
\reg{i\dirac}}\Big)^n\right)
\Bigg] \ , \cr
}\eqn\eqTRLOG
$$
The first term in the RHS is an irrelevant constant and the
second term vanishes as explained before.
Now we consider the $\Lambda \rightarrow \infty $ limit of this
quantity \eqTRLOG. Recall the propagator decomposition in
A picture:
$$
{1\over \reg{i\dirac}}=
{1\over {i\dirac}}
-{1\over 2}{1\over {i\dirac}-\Lambda }-{1\over 2}{1\over
{i\dirac}+\Lambda }\ ,
\eqn\eqPROPDEC
$$
and call the LHS regularized propagator, the first term
$1/i\dirac$ in the RHS unregularized propagator and the second
and third terms with masses $\pm \Lambda $ regulator propagators.
In the last term
$\Tr\big[-\sum _{n\geq 5}(1/n)(\calM/i\reg{\dirac})^n\big]$
in Eq.\eqTRLOG, we substitute \eqPROPDEC\ for each propagator
factor $1/i\reg{i\dirac}$ and then it becomes a sum of various
terms consisting of the unregularized and regulator propagators.
But, from dimension counting, each term is ultraviolet convergent
and therefore we can take the $\Lambda \rightarrow \infty $ limit
directly for the
loop integrands (\ie, inside the functional trace Tr).
Clearly then all the terms
containing the regulator propagators at least once drop out and
only the term
$\Tr\big[-\sum _{n\geq 5}(1/n)(\calM/i{\dirac})^n\big]$ survives.
If we add to the Tr operand of this term the quantity
$$
 \Ln\big[{i\dirac}\big]
 -{\calM\over {i\dirac}}
 -{1\over 2}\Big({\calM\over {i\dirac}}\Big)^2
 -{1\over 3}\Big({\calM\over {i\dirac}}\Big)^3
 -{1\over 4}\Big({\calM\over {i\dirac}}\Big)^4,
$$
then it reproduces the one-loop term
$\Tr\Ln\big[{i\dirac}-\calM\big]$ of the unregularized fermion.
We thus find that
$$
\eqalign{
\Tr\Ln\big[ \reg{i\dirac}-&\calM\big]
= \Tr\Ln\big[{i\dirac}-\calM\big]
+{i\over \Nc}\Gamma _{\rm count} +  O\Big({1\over \Lambda ^2}\Big)\ ,
\cr
\noalign{\vskip0.5cm}
{i\over \Nc}\Gamma _{\rm count} &\equiv
 - \bigg(
 {1\over 2}\Big({\calM\over  \reg{i\dirac}}\Big)^2
 +{1\over 3}\Big({\calM\over  \reg{i\dirac}}\Big)^3
 +{1\over 4}\Big({\calM\over  \reg{i\dirac}}\Big)^4
 \bigg) \cr
 &\qquad \ \ \  + \bigg(
 {1\over 2}\Big({\calM\over {i\dirac}}\Big)^2
 +{1\over 3}\Big({\calM\over {i\dirac}}\Big)^3
 +{1\over 4}\Big({\calM\over {i\dirac}}\Big)^4
 \bigg)
\ , \cr
}\eqn\eqTRLOGI
$$
up to an irrelevant constant.
[Precisely speaking, the last term and the second last term
$\Tr\big[-(1/4)(\calM/i\reg{\dirac})^4\big]$ in Eq.\eqTRLOG\
separately have an infrared divergence and
the discussion above is not so rigorous. But actually
a more careful argument can justify this expression \eqTRLOGI\
as is inferred from the fact that in this expression \eqTRLOGI\
the infrared divergence cancels between the terms
$\Tr\big[-(1/4)(\calM/i\reg{\dirac})^4\big]$ and
$\Tr\big[(1/4)(\calM/i{\dirac})^4\big]$.]

The Tr operation (loop integral) of this expression \eqTRLOGI\
as a whole of course gives a convergent quantity. But the
integral for each term separately is divergent.
In practice, however,
we can evaluate the integral for each term
separately if we use dimensional regularization.
The convergence as a whole implies that the pole terms $1/{\bar\epsilon
}$
appearing from each term should cancel eventually. We can use this
fact to check the calculations.

We can evaluate the `counterterm' $\Gamma _{\rm count}$
defined in Eq.\eqTRLOGI\ in a closed form if we discard
$O(1/\Lambda ^2)$ terms
since then only dimension 2 and 4 operators survive.
By using Eq.\eqTRLOGI,
the effective action \eqEFFACTION\ reads:
$$
\eqalign{
\Gamma  = \int d^4x &\ [\tree] \cr
&+{{\Nc}\over i}
\Tr\Ln[\,i\dirac-\calM\,] + \Gamma _{\rm count}
+ O\big({1\over \Lambda ^2}\big)\ .\cr
} \eqn\eqEFFACTIONI
$$
Performing a straightforward (but a bit laborious) calculation
we find that the counterterm $\Gamma _{\rm count}$ is given by
$$
\eqalign{
\Gamma _{count} = \ &\Gamma _{count}^{(2)}
+ \Gamma _{count}^{(3)} + \Gamma _{count}^{(4)} \ , \cr
\noalign{\vskip.5cm}
\Gamma _{count}^{(2)}\ \ \ \ \ & \cr
= {\Nc\over 16\pi ^2}\Bigg[
&2{\Lambda^2}
\tr({\Sigma^\dagger}\Sigma)
+(\VL-{4\over3})
\tr({\partial^\mu}{\Sigma^\dagger}{\partial_\mu}\Sigma)
-{1\over2}{\Lambda^2}\tr(R^\mu R_\mu+L^\mu L_\mu)
\cr
+\bigg\{
&-{1\over6}(\VL-{7\over6})
\tr\Big[{({\partial_\mu}{R_\nu}-{\partial_\nu}{R_\mu})}^2
\Big]
-{1\over6}\tr\Big[({\partial^\mu}{R_\mu})^2 \Big]
+ \big( R\rightarrow L \big) \bigg\}
\Bigg] \ ,\cr
\Gamma _{count}^{(3)} \ \ \ \ \ &\cr
= {\Nc\over 16\pi ^2}\Bigg[
&-i(\VL-{7\over3})
\tr\Big[{R^\mu}({\Sigma^\dagger}{\bpartial_\mu}\Sigma)
+{L^\mu}(\Sigma{\bpartial_\mu}{\Sigma^\dagger})\Big]
\cr
&+{i\over3}(\VL-2)
\tr\Big[({\partial^\mu}{R^\nu}-{\partial^\nu}{R^\mu})[{R_\mu},{R_\nu}]
+(R\rightarrow L)\Big]
\Bigg] \ ,\cr
\Gamma _{count}^{(4)}\ \ \ \ \ & \cr
= {\Nc\over 16\pi ^2}\Bigg[
&-(\VL-{17\over6})\tr\Big[({\Sigma^\dagger}\Sigma)^2\Big]
+(\VL-{7\over3})\tr\Big({R^2}{\Sigma^\dagger}\Sigma
+{L^2}\Sigma{\Sigma^\dagger}\Big)
\cr
&-2(\VL-{17\over6})\tr
\Big(\Sigma{R^\mu}{\Sigma^\dagger}{L_\mu}\Big) \cr
&+\bigg\{
{1\over6}(\VL-{29\over12})\tr\Big([{R_\mu},{R_\nu}]^2
\Big)
-{1\over12}\tr\Big[({R^\mu}{R_\mu})^2 \Big]
+ \big( R\rightarrow L \big) \bigg\}
\Bigg] \ ,\cr
} \eqn\eqCOUNT
$$
where $\Gamma _{count}^{(n)}$ with $n=2,3,4$ stand for the
contributions from
$(1/n)\Tr\Ln\big[-\big(\calM/\reg{i\dirac}\big)^n
+\big(\calM/{i\dirac}\big)^n \big]$, respectively, and
$\VL$ is the divergence factor
$\ln{\Lambda^2}-1/\bar\epsilon$
introduced in \eqDIVFAC.
It should be noted that the divergent terms proportional to $\VL$
in $\Gamma_{count}$ are combined to yield just the following
`gauge covariant' form:
$$
{\Nc\over 16\pi ^2}\VL
\Big[
\tr({D^\mu}{\Sigma^\dagger}{D_\mu}\Sigma)
-{1\over6}\tr({F^{R \mu \nu}}{F^R_{\mu \nu}}
+{F^{L \mu \nu}}{F^L_{\mu \nu}})
\Big]
\eqn\eqCOV
$$
with
$$
\eqalign{
{D_\mu}\Sigma &\equiv{\partial_\mu}\Sigma-i{L_\mu}\Sigma
+i\Sigma{R_\mu}\cr
{F^X_{\mu \nu}} &\equiv{\partial_\mu}{X_\nu}-{\partial_\nu}{X_\mu}
-i[{X_\mu},{X_\nu}] \qquad (X=R,L)\ . \cr
}\ee
$$
This `gauge covariant' form for the divergent parts
is a reflection of the fact that the
fermion one-loop term $\propto \Tr\Ln\big[\,\reg{i\dirac}-\calM\big]$
formally reduces to a gauge invariant form
$\Tr\Ln\big[
i\gamma ^\mu ( \partial _\mu -iR_\mu \calPR -iL_\mu \calPL )
          -\Sigma \calPR -\Sigma ^\dagger \calPL \big] $
in the $\Lambda \rightarrow \infty $ limit
as if $R_\mu $ and $L_\mu $ are the gauge fields of the local
$U(\nf)_{\rm R}\times U(\nf)_{\rm L}$ symmetry.
The finite parts, however, do not have such gauge invariance,
of course.

This form of the effective action \eqEFFACTIONI\ with counterterm
$\Gamma _{\rm count}$ is a very convenient one. The effect
of the cutoff by higher derivative term is now isolated in
the counterterm $\Gamma _{\rm count}$ and it is given explicitly
in \eqCOUNT. All we have to do then is to calculate the simple fermion
one-loop action $\Tr\Ln[\,i\dirac-\calM\,]$ with no cutoff by
using dimensional regularization. The divergences appearing there will
be automatically canceled by the contributions from the counterterm
$\Gamma _{\rm count}$.

Let us demonstrate the simplicity of this way of calculations
by taking an example --- the effective potential
$V(m)\equiv V\big( \Sigma \!=\! m{\bf1}_{\nf} \big)$.
The effective potential contribution from the unregularized fermion
one-loop is easily evaluated as
$$
\eqalign{
{1\over \nf}&V_{{\rm unreg.\,1\hbox{-}loop}}(m)
 =  - \Nc \int {d^np\over i(2\pi )^n} \tr\ln [ m-\pslash ] \cr
&\qquad  = - 2\Nc \int {d^np\over i(2\pi )^n} \ln\big[ m^2-p^2 ]
 = - {\Nc\over 16\pi ^2}
 m^4\Big( \VL - {3\over 2} + \ln{m^2\over \Lambda ^2} \Big) \ , \cr
}\eqn\eqEFFECTIVEPOTI
$$
while the contribution from the counterterm is found by
substituting $\Sigma =$ $m{\bf1}_{\nf}$ into \eqCOUNT\
to be
$$
{1\over \nf}V_{\rm count}(m)
= - {\Nc\over 16\pi ^2}
 \Big( 2m^2\Lambda ^2 - m^4 \big( \VL - {17\over 6} \big) \Big) \ .
\ee
$$
We see the divergence $\propto \VL$ in \eqEFFECTIVEPOTI\ is actually
canceled by this counterterm. Adding the tree contribution
$\nf(m^2\Lambda ^2/2g_\S^2)$ also, we have
$$
{1\over \nf}V(m) = {\Lambda ^2\over 2g_\S^2}m^2
 - {\Nc\over 16\pi ^2} \Big( 2m^2\Lambda ^2
+ m^4\big( \ln{m^2\over \Lambda ^2} + {4\over 3} \big) \Big) \ .
\eqn\eqPOTII
$$
The validity of this expression is of course limited in the
region $m^2/\Lambda ^2 \ll  1$.  We draw in Fig.3 pictures of
this potential and the `exact' potential
\eqPOTPOTPOT\ in Sect.3, for comparison.
\midinsert
\checkex{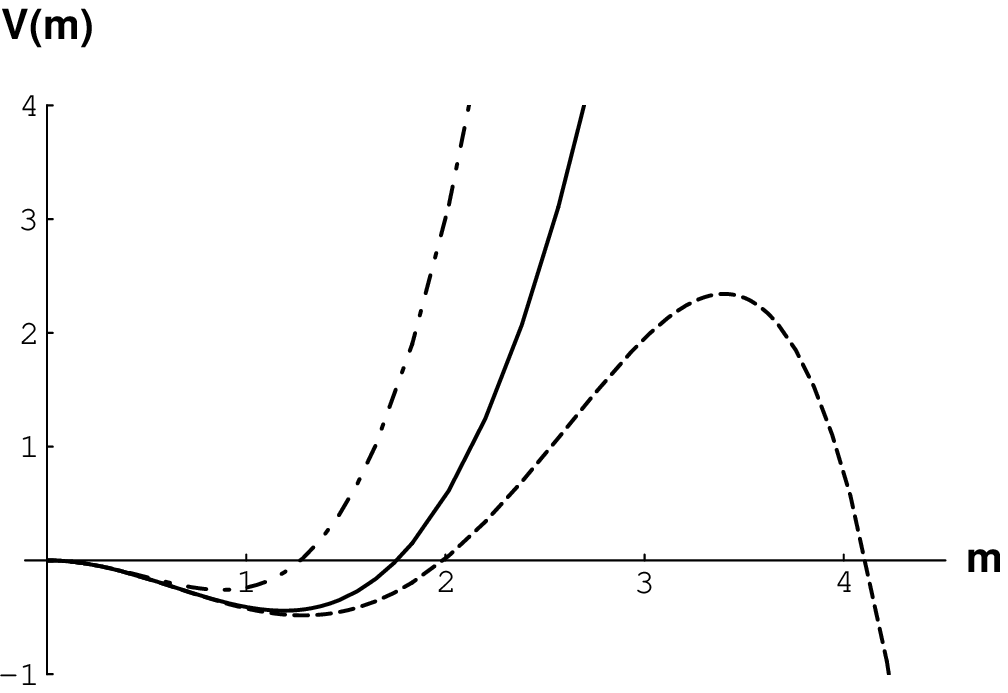}
\iffigureexists
    \immediate\write20{>Embedding an epsf Figure (effpot.eps).}
\epsfxsize=10cm
\centerline{\epsfbox{effpot.eps}}
\vskip -1cm
\else
    \immediate\write20{>Cannot embed a epsf Figure (effpot.eps).}
\fbox{Figure 3}
\fi
\captionbox{3}
{Effective potentials for the case $g_\S=1.0193g_\S^{\rm cr}$ above
the critical coupling $g_\S^{\rm cr}=2\pi/\sqrt{\Nc}$.
Vertical and horizontal axes give scaled ones:
$V(m)\times (16\pi^2\cdot 10^3/\Nc\Lambda^4)$ and
$m\times(10/\Lambda)$. Broken line denotes the
approximate one (5.10), solid line the `exact' one (3.10) and
dotted-broken line the one (5.11) of the cutoff theory.}
\endinsert
It may also be of some interest to compare this potential \eqPOTII\
with the usual effective potential obtained by the simple cutoff of the
loop momentum which reads
$$
{1\over \nf}V(m) = {\Lambda ^2\over 2g_\S^2}m^2
 - {\Nc\over 16\pi ^2} \Big( m^2\Lambda ^2 + \Lambda
^4\ln\big(1+{m^2\over \Lambda ^2}\big)
 - m^4\ln\big(1+{\Lambda ^2\over m^2}\big) \Big) \ .
\eqn\eqPOTII
$$
We have also drawn this potential function in Fig.3.
[Its deviation from the present `exact'
one (3.10) in the large $m$ region in Fig.3 reflects the difference in
the meaning of the cutoff $\Lambda$ in both theories.]

\chapter{Conclusion}

In this paper we have defined the NJL model by introducing higher
derivative fermion kinetic term. We have clarified some basic aspects
of the theory such as quantization, current operators \etc, and
developed two calculation methods which make the evaluation of the
diagrams in this higher derivative system no more difficult than in
the usual first order derivative case.

We have emphasized that the present formulation of the NJL model
suffers from no ambiguities in the loop momentum assignments and keeps
the important chiral and gauge symmetries. We explicitly demonstrated
this by calculating the $\pi_0 \rightarrow 2\gamma$ amplitude and
confirming that the low energy theorem holds.

We restricted ourselves in this paper to the system with exact chiral
symmetry. But there will be no problem in extending the present
formalism to the cases where the fermions have explicit chiral
symmetry breaking masses. When a fermion has an explicit mass $M$, we
think it best to take the kinetic term in the form:
$$
\bar\psi \big(i\dirac-M\big)
\left( 1+ { \dirac\dirac + M^2 \over \Lambda^2} \right)\psi \ .
\ee
$$
Indeed, then the fermion really has mass $M$ in the absence of
interactions and satisfies correct normalization. The fermion momentum
is effectively cut off as $|p^2-M^2|\lsim \Lambda^2$ around the mass
shell. This form will be suitable also for describing heavy quarks
which are much studied recently in connection with heavy quark
symmetry\rlap.\refmark{\BH}

\vskip 1cm

{}~
\acknowledge
{The authors would like to thank K.-I.~Aoki, M.~Bando, Y.~Kikukawa,
M.~Harada, T.~Maskawa, M.~Sasaki, H.~Tagoshi, T.~Tanaka and H.~Sato
for various discussions and encouragements.
T.K.\ is supported in part by the Grant-in-Aid for Scientific Research
(\#06640387) from the Ministry of Education, Science and Culture.}

\APPENDIX{A}{\  Noether Current for a generic Higher Derivative
System}

We consider a generic system whose action contains
arbitrary order derivatives:
$$
S(\phi ) = \int {d^4}x \, {\cal L}\big(\phi ,\ \partial _\mu \phi ,\
  \partial _{\mu \nu }\phi , \ \partial _{\mu \nu \rho }\phi , \ \cdots
\big)\ ,
\ee
$$
where $\phi $ stands for a collection of fields
(whose index is suppressed) and we use abbreviations like
$$
\eqalign{
\partial _{\mu _1\mu _2\cdots \mu _n} \phi
&\equiv  \partial _{\mu _1}\partial _{\mu _2}\cdots \partial _{\mu
_n}\phi \ , \cr
\delta {\cal L}^{\,;\mu _1\mu _2\cdots \mu _n}
&\equiv  {\partial {\cal L}\over \partial  (\partial _{\mu _1\mu
_2\cdots \mu _n}\phi ) }
\Big|_{\rm weight\,1}  \ .
\cr}\ee
$$
The suffix `weight 1' in the latter means that we keep always the
weight to be one irrespectively of whether the $n$ indices
$\mu _1, \mu _2, \cdots , \mu _n$ take the same values or not;
namely, for the case
${\cal L}= a^{\mu \nu }\partial _{\mu \nu }\phi $, for instance, \
$\partial {\cal L}/\partial (\partial _{11}\phi )= a^{11}$ and
$\partial {\cal L}/\partial (\partial _{12}\phi )= a^{12}+a^{21}$,  but
$\partial {\cal L}/\partial (\partial _{\mu \nu }\phi )
\big|_{\rm weight\, 1}= (a^{\mu \nu }+a^{\nu \mu })/2!$
\ always.
The functional derivative of the action $S$ with respect to $\phi$ is
given by
$$
 {{\delta S}\over{\delta \phi}}
={{\partial {\cal L}}\over{\partial \phi}}
-{\partial_\mu}(\delta{\cal L}^{\,;\mu })+{\partial_{\mu \nu}}
(\delta{{\cal L}^{\,; \mu \nu}})
-{\partial_{\mu \nu \rho}}(\delta {{\cal L}^{\,;\mu \nu \rho}})
  +\cdots \ ,
\ee
$$
and the Euler-Lagrange equation of motion is written as
$\delta S/\delta \phi =0$.
If we perform an infinitesimal transformation
$\phi  \rightarrow \phi  + \delta \phi $,
the Lagrangian ${\cal L}$ changes as
$$
\eqalign{
\delta {\cal L}&=
{\partial {\cal L}\over \partial \phi }\delta \phi
+(\delta {\cal L}^{\,;\mu })\partial _\mu \delta \phi
+(\delta {\cal L}^{\,;\mu \nu })\partial _{\mu \nu }\delta \phi
+(\delta {\cal L}^{\,;\mu \nu \rho })\partial _{\mu \nu \rho }\delta
\phi
+ \cdots  \cr
&=
{\delta S\over \delta \phi }\delta \phi
+\big[(\delta {\cal L}^{\,;\mu })\partial _\mu \delta \phi
+\partial _\mu (\delta {\cal L}^{\,;\mu })\cdot \delta \phi \big]
+\big[(\delta {\cal L}^{\,;\mu \nu })\partial _{\mu \nu }\delta \phi
-\partial _{\mu \nu }(\delta {\cal L}^{\,;\mu \nu })\cdot \delta \phi
\big] \cr
& \ \ \ \ \ \ \ \qquad
+\big[(\delta {\cal L}^{\,;\mu \nu \rho })\partial _{\mu \nu \rho
}\delta \phi
+\partial _{\mu \nu \rho }(\delta {\cal L}^{\,;\mu \nu \rho })\cdot
\delta \phi \big]
+ \cdots  \ .\cr
}\eqn\eqDELTALAG
$$
To rewrite this, we introduce a generalized `both-side' derivative
defined by
$$
\eqalign{
F\,{\mathop{\partial }^{\leftrightarrow }}_{\mu _1\mu _2\cdots \mu _n}
\,G
\ \equiv \ & F\partial _{\mu _1\mu _2\cdots \mu _n}G
- \partial _{\mu _1}F\cdot \partial _{\mu _2\cdots \mu _n}G \cr
 & + \partial _{\mu _1\mu _2}F\cdot \partial _{\mu _3\cdots \mu _n}G
  - \cdots  +(-)^n\partial _{\mu _1\mu _2\cdots \mu _n}F\cdot G \ , \cr
}\ee
$$
for arbitrary two functions $F$ and $G$.
This derivative is no longer symmetric under permutation of
the indices but enjoys an identity
$$
\partial _\mu \big[ F^{\mu \alpha _1\cdots \alpha _n}
{\mathop{\partial }^{\leftrightarrow }}_{\alpha _1\cdots \alpha _n}
\,G\big]
= F^{\mu \alpha _1\cdots \alpha _n}\partial _{\mu \alpha _1\cdots
\alpha _n}G
+ (-)^n\partial _{\mu \alpha _1\cdots \alpha _n}F^{\mu \alpha _1\cdots
\alpha _n}\cdot G
\ee
$$
for functions $F^{\mu \alpha _1\cdots \alpha _n}$
totally symmetric with respect to the $n+1$ indices
$\mu , \alpha _1,$ $\cdots , \alpha _n$.  Applying this identity we can
write
\eqDELTALAG\ as
$$
\eqalign{
\delta {\cal L} &=
{\delta S\over \delta \phi }\delta \phi
+\partial _\mu \big[(\delta {\cal L}^{\,;\mu })\delta \phi \big]
+\partial _\mu \big[(\delta {\cal L}^{\,;\mu \nu })
{\mathop{\partial }^{\leftrightarrow }}_{\nu }\delta \phi \big]
+\partial _\mu \big[(\delta {\cal L}^{\,;\mu \nu \rho })
{\mathop{\partial }^{\leftrightarrow }}_{\nu \rho }\delta \phi \big]
+ \cdots  \cr
&=
{\delta S\over \delta \phi }\delta \phi
+\partial _\mu \left[\sum _{n=0}^\infty
(\delta {\cal L}^{\,;\mu \alpha _1\cdots \alpha _n})
{\mathop{\partial }^{\leftrightarrow }}_{\alpha _1\cdots \alpha
_n}\delta \phi \right]\ .\cr
}\ee
$$

If the transformation $\phi \rightarrow \phi +\delta \phi $
with $\delta \phi \equiv \varepsilon ^a\trans$
($\varepsilon ^a$: $x$-independent transformation parameters)
leaves the lagrangian invariant, $\delta {\cal L}=0$, then we obtain
from this an identity:
$$
\eqalign{
&\partial _\mu j_a^\mu  = -{\delta S\over \delta \phi }\trans \ , \cr
&\quad  j_a^\mu  \equiv  \sum _{n=0}^\infty
(\delta {\cal L}^{\,;\mu \alpha _1\cdots \alpha _n})
{\mathop{\partial }^{\leftrightarrow }}_{\alpha _1\cdots \alpha
_n}\trans \ .\cr
}\ee
$$
This $j_a^\mu $ gives a generalized Noether current and is seen to be
conserved if the Euler-Lagrange equation
$\delta S/\delta \phi =0$ is satisfied.

For completeness we show here that this generalized Noether current
coincides with the source current of the gauge field if it is
introduced by the usual covariantization procedure replacing the
derivative by a covariant one:
$\partial _\mu \rightarrow D_\mu \equiv \partial _\mu -iA^a_\mu T^a$
for the case $\hat\delta _a\phi =iT^a\phi $ with a certain
representation
matrix $T^a$. Then noting
$$
\eqalign{
&\partial _{\mu _1\mu _2\cdots \mu _n}\phi  \ \rightarrow  \ \cr
&\ D_{\mu _1}D_{\mu _2}\cdots D_{\mu _n}\phi  =
\partial _{\mu _1\mu _2\cdots \mu _n}\phi  -
\sum _{k=1}^n \partial _{\mu _1\cdots \mu _{k-1}}
\big(A_{\mu _k}^a\partial _{\mu _{k+1}\cdots \mu _n}\hat\delta _a\phi
\big)
+ O(A^2) \ , \cr
}\ee
$$
we find the term linear in the gauge field in the covariantized
lagrangian ${\cal L}_{\rm cov}(\phi ,A)$ is given by
$$
\eqalign{
&{\cal L}_{\rm cov}(\phi ,A)\big\vert_{A {\rm \hbox{-}linear} }
= -\sum _{n=0}^\infty
\sum _{k=0}^n (\delta {\cal L}^{\,;\mu \alpha _1\cdots \alpha
_n})\partial _{\alpha _1\cdots \alpha _k}
\big(A_{\mu }^a\partial _{\alpha _{k+1}\cdots \alpha _n}\hat\delta
_a\phi \big) \cr
&\quad = -A_{\mu }^a\,\sum _{n=0}^\infty  \left[
\sum _{k=0}^n (-)^k\partial _{\alpha _1\cdots \alpha _k}(\delta {\cal
L}^{\,;\mu \alpha _1\cdots \alpha _n})\cdot
\partial _{\alpha _{k+1}\cdots \alpha _n}\hat\delta _a\phi \right] +
\big(\hbox{tot. der.} \big) \ , \cr
}\ee
$$
where we have performed a `partial integration' in going to the
second line and (tot. der.) denotes a total derivative term
appearing then.
We note that the quantity multiplied by $A_\mu ^a$ in the first term
is just identical with the above Noether current and so we have
shown
$$
-{\delta S_{\rm cov}[\phi ,A]\over \delta A_\mu ^a}\Big\vert_{A=0}
=  \ j^\mu _a \ .
\ee
$$

\refout

\end